%% file: main.tex
\documentclass[10pt,twocolumn,letterpaper]{article}
\usepackage[pagenumbers]{cvpr} 
\usepackage[accsupp]{axessibility}

\input{preamble}
\usepackage{bbm}

\definecolor{cvprblue}{rgb}{0.21,0.49,0.74}
\input{shortcut}

\author{Samuel Young\textsuperscript{\mdseries1,2} \authorskip 
Kazuhiro Terao\textsuperscript{\mdseries2}
 \\ \\
\textsuperscript{1}Stanford University
\institutionskip
\textsuperscript{2}SLAC National Accelerator Laboratory \\ 
{\tt\small \url{https://github.com/DeepLearnPhysics/Panda}}
}

\begin{document}
\title{Panda: Self-distillation of Reusable Sensor-level Representations \\for High Energy Physics}

\twocolumn[{%
\renewcommand\twocolumn[1][]{#1}%
\vspace{-17mm}
\maketitle
\vspace{-12mm}
\begin{center}
    \captionsetup{type=figure}
    \includegraphics[height=0.65\linewidth]{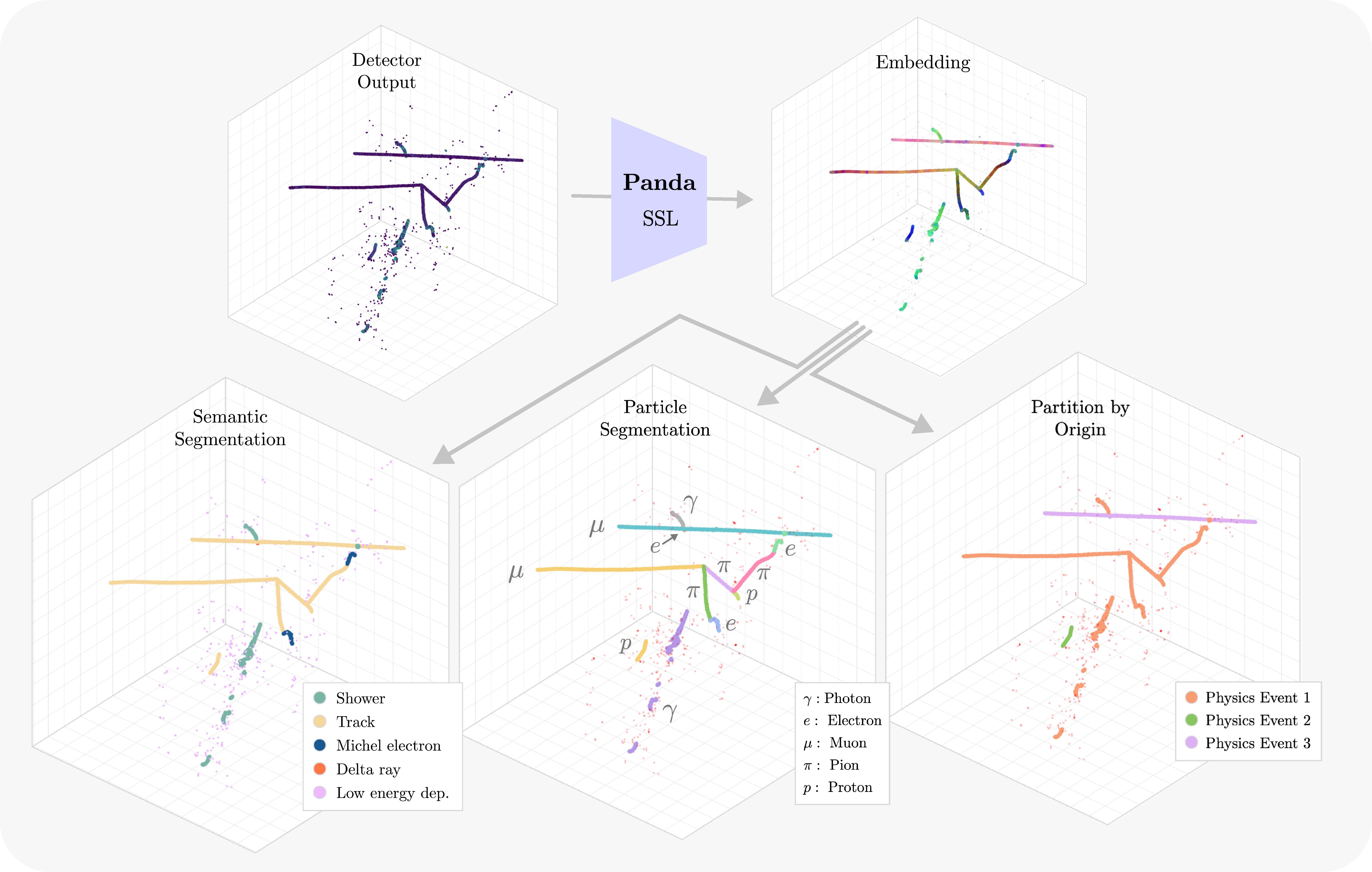}
    \vspace{-4mm}
    \captionof{figure}{
    \textbf{Panda overview.}
    Raw charge depositions corresponding to particle trajectories recorded by a time projection chamber (TPC) (top left) are passed through a point-native hierarchical encoder pre-trained via self-distillation (\emph{Panda}) to produce a shared embedding (top right).
    The same pretrained features are used for three downstream tasks with lightweight heads (bottom): semantic segmentation of geometric motifs; particle-level panoptic segmentation with per-particle masks and IDs (\(\gamma,e,\mu,\pi,p\)); and interaction-level partitioning that groups causally related particles. A single sensor-level backbone supports all tasks without detector-specific heuristics.
    }\label{fig:teaser}
    \vspace{-2mm}
\end{center}%
}]

\input{secs/0_abstract}    
\input{secs/1_introduction}
\input{secs/2_related_work}

\input{secs/3_method}
\input{secs/4_results}
\input{secs/5_conclusion}
\input{secs/7_acknow}

{
    \small
    \bibliographystyle{ieeenat_fullname}
    \bibliography{main.bib}
}

\appendix
\input{secs/6_suppl}

\end{document}

%% file: preamble.tex
\usepackage{pgf}

%% file: shortcut.tex
\usepackage{lipsum}
\usepackage{array}
\usepackage{times}
\usepackage{epsfig}
\usepackage{graphicx}
\usepackage{float}
\usepackage{wrapfig}
\usepackage{amsmath,amssymb,amsthm}
\usepackage{algorithm,algorithmicx,algpseudocode}
\usepackage{bm,xspace}
\usepackage{comment}
\usepackage{multirow}
\usepackage{balance}
\usepackage{url}
\usepackage{booktabs}
\usepackage{etoolbox,siunitx}
\usepackage{calc}
\usepackage{pifont,hologo}
\usepackage{color}
\usepackage{adjustbox}
\usepackage{amsmath}
\usepackage{enumitem}
\usepackage{bbding}  %
\usepackage[normalem]{ulem}  %
\usepackage{contour}
\usepackage[table]{xcolor}
\usepackage{soul}%
\usepackage{tocloft} %
\usepackage{dsfont}
\usepackage{etoc} %
\PassOptionsToPackage{dvipsnames}{xcolor}
\usepackage{algorithm}
\usepackage{algpseudocode}
\usepackage{listings}
\usepackage[pagebackref,breaklinks,colorlinks,linkcolor=link,citecolor=cite,breaklinks=true]{hyperref}
\PassOptionsToPackage{square,numbers,comma,sort}{natbib}

\newcommand{\authorskip}{\hspace{2.5mm}}
\newcommand{\institutionskip}{\hspace{5.0mm}}

\definecolor{darkblue}{HTML}{35394B}
\definecolor{blue}{HTML}{004bb3}
\definecolor{red}{HTML}{cc1100}
\definecolor{orange}{HTML}{cc7700}
\definecolor{gray}{HTML}{efefef}
\definecolor{darkgreen}{HTML}{228B22}
\definecolor{darkgray}{HTML}{808080}
\definecolor{lightpurple}{HTML}{a56dba}
\definecolor{white}{HTML}{ffffff}
\definecolor{C0}{HTML}{1f77b4}
\definecolor{C1}{HTML}{ff7f0e}
\definecolor{C2}{HTML}{2ca02c}
\definecolor{cite}{HTML}{3270b5}
\definecolor{link}{HTML}{b53532}
\definecolor{link}{HTML}{cc1100}
\definecolor{scratch}{HTML}{001219}
\definecolor{pretrain}{HTML}{0a9396}

\newcommand{\sota}{\textcolor{lightpurple}{$\mathbf{\circ}$\,}}
\newcommand{\thiswork}{\textcolor{lightpurple}{$\bullet$\,}}
\newcommand{\otherwork}{\textcolor{white}{$\bullet$\,}}

\contourlength{0.8pt}

\renewcommand{\eqref}[1]{Eq.~\ref{#1}}

\newcolumntype{x}[1]{>{\centering\arraybackslash}p{#1}}
\newcolumntype{y}[1]{>{\raggedright\arraybackslash}p{#1}}
\newcolumntype{z}[1]{>{\raggedleft\arraybackslash}p{#1}}
\newcommand{\tablestyle}[2]{\setlength{\tabcolsep}{#1}\renewcommand{\arraystretch}{#2}\centering\footnotesize}

\DeclareMathSymbol{@}{\mathord}{letters}{"3B}

\makeatletter
\DeclareRobustCommand\onedot{\futurelet\@let@token\@onedot}
\def\@onedot{\ifx\@let@token.\else.\null\fi\xspace}

\newcommand*{\Rom}[1]{\expandafter\@slowromancap\romannumeral #1@}
\newcommand*{\rom}[1]{\expandafter\romannumeral #1}

\def\1{\bm{1}}

\DeclareMathAlphabet{\mathsfit}{\encodingdefault}{\sfdefault}{m}{sl}
\SetMathAlphabet{\mathsfit}{bold}{\encodingdefault}{\sfdefault}{bx}{n}

\let\originalleft\left
\let\originalright\right
\renewcommand{\left}{\mathopen{}\mathclose\bgroup\originalleft}
\renewcommand{\right}{\aftergroup\egroup\originalright}

%% file: secs/0_abstract.tex
\begin{abstract}
Liquid argon time projection chambers (LArTPCs) provide dense, high-fidelity 3D measurements of particle interactions and underpin current and future neutrino and rare-event experiments. Physics reconstruction typically relies on complex detector-specific pipelines that use tens of hand-engineered pattern recognition algorithms or cascades of task-specific neural networks that require extensive, labeled simulation that requires a careful, time-consuming calibration process. We introduce \textbf{Panda}, a model that learns reusable sensor-level representations directly from raw unlabeled LArTPC data. Panda couples a hierarchical sparse 3D encoder with a multi-view, prototype-based self-distillation objective. On a simulated dataset, Panda substantially improves label efficiency and reconstruction quality, beating the previous state-of-the-art semantic segmentation model with 1,000$\times$ fewer labels. We also show that a single set-prediction head 1/20th the size of the backbone with no physical priors trained on frozen outputs from Panda can result in particle identification that is comparable with state-of-the-art (SOTA) reconstruction tools. Full fine-tuning further improves performance across all tasks.
\end{abstract}

%% file: secs/1_introduction.tex
\section{Introduction}
\label{sec:intro}

Particle physics seeks precise measurements of fundamental particles and their interactions. A central tool in recording the behavior of particles is the liquid argon time projection chamber (LArTPC)~\cite{rubbia1977liquid}, which is deployed in current and upcoming experiments that will enable high-precision measurements of neutrino oscillations~\cite{Acciarri_2017,Abratenko2023,sbn2015,Abi2020}, detection and physics modeling of Supernovae~\cite{Abi_2021_sn,PhysRevD.107.112012,PhysRevD.111.092006}, solar and atmospheric neutrinos~\cite{PhysRevLett.123.081801,PhysRevLett.123.131803,Kelly_2022,htfm-tbdq}, and a spectrum of physics beyond the Standard Model~\cite{Abi2020,Abi_2021,PhysRevD.101.052001, ARAMAKI2020107,shutt2024gampix,PhysRevD.103.095012,Kudryavtsev_2016}. LArTPCs are the central technology of the upcoming DUNE experiment~\cite{Abi2020}, the United States' flagship particle physics project. These detectors are high-resolution 3D imagers~\cite{app11062455}, providing millimeter-scale sampling of charge depositions over meter-scale volumes (Fig.~\ref{fig:teaser}, top left).

Particles recorded in LArTPC event data exhibit visually distinct patterns. Some appear as long, nearly straight tracks; others form broad, diffuse showers or short, dense clusters. A single particle type can take multiple forms: for example, electrons may appear as extended showers, small stub-like features along other tracks (Delta rays), or short tracks (Michel electrons) produced at the end of a muon as it decays (Fig. \ref{fig:teaser}, bottom left). This inter-class diversity and intra-class multimodality parallels natural image categories, such as animals and their different breeds and poses.

Whereas typical 3D perception benchmarks such as LiDAR scans or structure from motion (SfM) reconstructions emphasize robustness under sparsity and noise as they are limited by the instrumentation in which these images are taken, LArTPC data provide dense, high-fidelity measurements in which essentially every active pixel carries semantic and physical information; the objective is exhaustive, pixel-level understanding. The mapping from observed geometry to underlying physics is direct and repeatable, with relatively low semantic ambiguity at the pattern level, which expert physicists can intuitively identify.\footnote{This is not always the case, however. For example, non-interacting or uncontained pions can often look like muons, and photon- and electron-initiated electromagnetic (EM) showers can be indistinguishable. See Sec.~\ref{sec:results}.} We argue that these attributes align well with the objectives of modern self-supervised learning (SSL), which learn invariances and cluster structure from recurring geometric and textural motifs present at scale, without task-specific labels.

\begin{figure}
    \centering
    \includegraphics[width=1.0\linewidth]{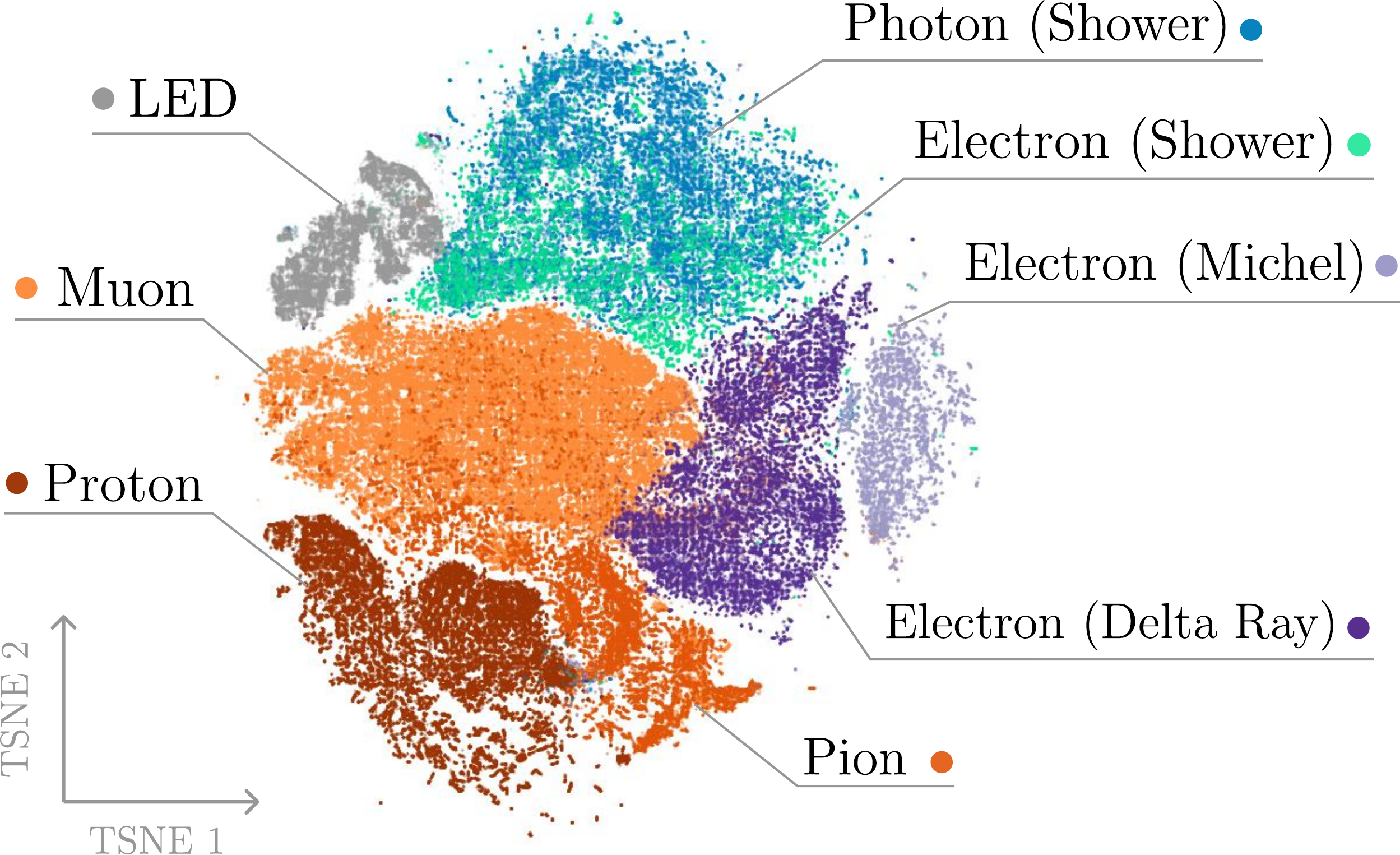}
    \caption{\textbf{t-SNE visualization of Panda embeddings.} We visualize per-point embeddings from 1,000 images via t-SNE. Clear structure that corresponds to the inter-class diversity and intra-class multi-modality within LArTPC images is exhibited.}
    \vspace{-5mm}
    \label{fig:tsne}
\end{figure}

Despite this, reconstruction currently relies on detector-specific chains such as Pandora~\cite{Acciarri2018} and SPINE~\cite{drielsma2021scalableendtoenddeeplearningbaseddata}; these are large collections of hand-crafted algorithms (Pandora) and task-specific networks (SPINE) that lack flexibility, tied to simulation, and difficult to transfer. This motivates a unified sensor-level backbone learned directly from raw data.

Recent progress in prototype-based soft clustering and self-distillation \cite{oquab2024dinov2learningrobustvisual, wu2025sonataselfsupervisedlearningreliable} has produced strong foundation models for natural images \cite{caron2021emerging, oquab2024dinov2learningrobustvisual, swav, zhou2022ibotimagebertpretraining} and 3D scenes \cite{wu2025sonataselfsupervisedlearningreliable, concerto}, where teacher–student architectures with prototype assignments and invariance constraints learn reusable features across tasks and datasets. In parallel, first attempts (PoLAr-MAE~\cite{polarmae}) at self-supervised learning on LArTPC data have demonstrated that masked modeling on 3D charge clouds can produce latent spaces aligned with physically meaningful motifs and improve low-label semantic segmentation performance over purely supervised baselines. These results indicate that sensor-level SSL can extract physically relevant structure. However, PoLAr-MAE operates on a coarse, fixed-radius tokenization of the 3D volume and reconstruct aggregated patches. Empirically, this constraint limits performance in the high-data regime and restrict sensitivity to subtle features such as short prongs, contiguous boundaries between particles, and rare topologies.

This work introduces \textbf{Panda}, a \textbf{po}int \textbf{a}ttention a\textbf{n}d \textbf{d}istill-\textbf{a}tion model for LArTPC data that targets reusable sensor-level representations through two design choices aligned with the structure of the detector and the demands of precise reconstruction. First, we use a clustering-based self-distillation objective: a teacher network maps augmented sets of particle trajectories to assignments over a set of learned prototypes, and a student network is trained to predict consistent prototype distributions across strong augmentations, without labels. Second, we eschew fixed-resolution embeddings, instead employing a point-native hierarchical encoder \cite{wu2024pointtransformerv3simpler} that operates directly on 3D charge clouds across multiple scales. This results in a generally reusable and extensible backbone that outputs per-point features~(Fig. \ref{fig:teaser}, top right).

We evaluate Panda on pixel-level semantic segmentation and particle- and interaction-level reconstruction tasks~(Fig. \ref{fig:teaser}, bottom row) for LArTPC data across label and trainable parameter sizes. The learned representations yield clearly separable clusters for key particle classes and geometric motifs (Fig.~\ref{fig:tsne}), indicating that the self-distilled embeddings align with physically meaningful structure without supervision. In data-scarce environments, pre-training with Panda followed by lightweight adaptation substantially improves semantic segmentation performance over both prior self-supervised methods (+19\% mean $F_1$), and the previous state-of-the-art fully supervised UResNet (+47\% mean $F_1$). In full fine-tuning settings, Panda achieves a new state-of-the-art on semantic segmentation of LArTPC data with 1,000$\times$ less labels. Fine-tuning on all data results a mean $F_1$ improvement of 3.2\% over the state-of-the-art and 0.6\% over training the same model from scratch. We show that a small set-prediction head attached to frozen Panda representations can result in particle identification performance comparable to that of SPINE~\cite{drielsma2021scalableendtoenddeeplearningbaseddata}.
These results establish self-distilled, point-native sensor-level pre-training as an effective route to robust and transferable representations for high energy physics detectors.

In sum, this paper makes the following contributions:

\begin{itemize}

\item Panda, a self-distillation pre-training framework for LArTPC data that combines prototype-based clustering under physically plausible augmentations with a point-native hierarchical encoder for 3D images of particle trajectories.

\item Empirical evidence that Panda learns sensor-level embeddings where particle types and geometric motifs are clearly separable without labels, supporting effective downstream reconstruction.

\item Systematic evaluation of semantic segmentation performance across low-label and full-label regimes, showing substantial gains over prior self-supervised approaches and supervised baselines, including a 3.2\% mean F1 improvement in the high-data regime.

\item Demonstration that a cascade of hand-crafted task-specific pattern recognition algorithms or neural networks are not necessary for competitive particle identification performance.

\item Evidence that one can achieve strong performance in any of the three aforementioned downstream tasks within $10-30\times$ less iterations compared to training the same model from scratch.

\end{itemize}

\noindent We also make our data and code available for general use.

%% file: secs/2_related_work.tex
\section{Related Work}
\paragraph{Self-supervised learning for images and 3D point clouds.} Recent advances in self-supervised learning across language, images, and 3D point clouds have established a new paradigm: pre-train large models on large unlabeled datasets, then reuse the backbone as a starting point for downstream tasks \cite{caron2021emerging,oquab2024dinov2learningrobustvisual}. This paradigm provides strong, reusable generic features that speed up prototyping and, in turn, scientific progress, often beating the state-of-the-art. Three families of objectives dominate large-scale image/3D pre-training: \emph{self-distillation} (e.g., DINO~\cite{caron2021emerging,oquab2024dinov2learningrobustvisual}, BYOL~\cite{grill2020byol}, iBOT~\cite{zhou2022ibotimagebertpretraining}, SwAV~\cite{swav}, and recent 3D adaptations~\cite{wu2025sonataselfsupervisedlearningreliable,concerto}), \emph{contrastive methods} (e.g., SimCLR~\cite{chen2020simple}, MoCo~\cite{he2020momentum}, PointContrast~\cite{xie2020pointcontrastunsupervisedpretraining3d}), and \emph{masked autoencoding} (e.g., MAE~\cite{he2021maskedautoencodersscalablevision}, MViT~\cite{mvit, mvitv2}, Hiera~\cite{ryali2023hierahierarchicalvisiontransformer}, with point-token variants~\cite{pang2022maskedautoencoderspointcloud,yu2022pointbertpretraining3dpoint}). Our images lend themselves to categorical recognition, so we explicitly test whether self-distillation is an effective pre-training paradigm for 3D images of particle trajectories. To do this, we ask whether a single pre-trained feature extractor can be reused across disparate tasks (Fig~\ref{fig:teaser}, bottom three images) with fast adaptation.

\paragraph{Foundation models in high energy physics.}
Much progress towards a ``foundation model'' for high energy physics has contributed by the particle collider community. These efforts typically operate on \emph{reconstructed} objects (e.g., jets, particle four-vectors). Recent self-supervised directions include masked particle modeling (MPM)~\cite{golling2024maskedparticlemodelingsets,leigh2024tokenizationneededmaskedparticle} on unordered particle sets, which masks constituents and trains set-equivariant encoders to recover discretized tokens; re-simulation–based SSL (RS3L)~\cite{r3sl,rieck2025selfsupervisedlearningstrategiesjet}, which treats physics-preserving variations of the same event as positive pairs; and self-distillation objectives designed for jets that enforce invariances expected from renormalization-group arguments (RINO)~\cite{hao2025rinorenormalizationgroupinvariance}. In parallel, supervised multi-task pretraining with massive cheaply simulated datasets has been explored to produce broadly useful jet feature extractors (e.g., joint classification/generative heads trained end-to-end)~\cite{Mikuni_2025,bhimji2025omnilearnedfoundationmodelframework}.

By contrast, the (LAr)TPC community works directly at the \emph{sensor level} with high-resolution, effectively noiseless 3D charge measurements. Early steps toward sensor-level SSL include SimCLR~\cite{chen2020simple}-like contrastive pretraining~\cite{wilkinson2025contrastivelearningrobustrepresentations} masked modeling~\cite{polarmae}, and autoregressive modeling~\cite{park2025fm4npp} on voxelized TPC point clouds; these confirm that unsupervised objectives can surface physically meaningful motifs within representations from raw detector measurements, but it has yet to be shown whether these networks are transferable or powerful enough to reach the production-level performance in reconstruction tasks. Our work advances this sensor-level line by pre-training on raw 3D charge clouds with a \emph{self-distillation} objective~\cite{wu2025sonataselfsupervisedlearningreliable} and a point-native hierarchical encoder~\cite{wu2024pointtransformerv3simpler}.

\begin{figure*}[ht]
    \centering
    \includegraphics[width=1.0\linewidth]{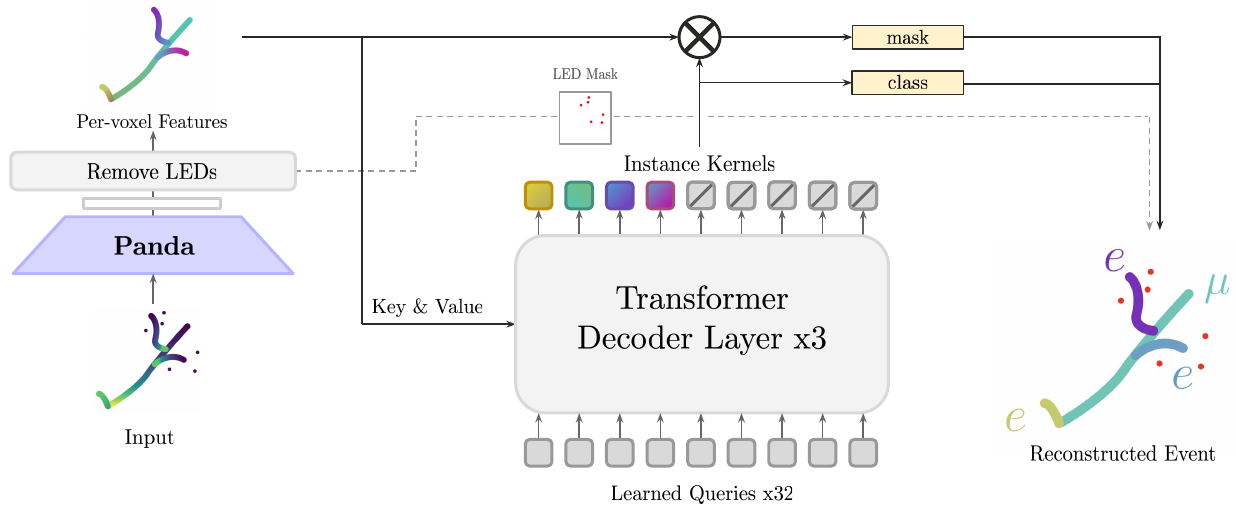}
    \vspace{-5mm}
    \caption{\textbf{Panoptic segmentation architecture.} We adopt a Mask2Former~\cite{cheng2021perpixelclassificationneedsemantic}-like set prediction module for the panoptic segmentation task. The image is first encoded with the pre-trained Panda backbone. A small MLP removes low energy deposits from the image; the remaining point embeddings are fed into the cross attention mechanism of a small three-layer transformer that decoded learned queries into a set of instance kernels. Masks for each kernel are created by taking the sigmoid of the dot product between the kernel and per-point features. Classification of each kernel into one of five particle types as well as a no-object class is done with a small MLP. The full reconstructed image, with LEDs added back in, is then created using this set of predictions.}
    \label{fig:insseg}
    \vspace{-3mm}
\end{figure*}

\paragraph{Reconstruction of particles and their interactions in TPCs} 
LArTPC reconstruction has largely followed two paths. The first is a hand-engineered, detector-specific chain: Pandora~\cite{Acciarri2018} and WireCell~\cite{Abratenko_2022}, the current modus operandi for event reconstruction in DUNE, stitches together dozens of pattern-recognition algorithms tuned to particular running conditions and topologies. The second, SPINE~\cite{drielsma2021scalableendtoenddeeplearningbaseddata}, replaces these algorithms with a cascade of task-specific neural networks (voxel-level semantic segmentation, dense clustering, primary identification, and ultimately particle/interaction assembly). These systems work, but they are complex to maintain, hungry for simulated labels, and difficult to calibrate to real data; as a cascade, errors in earlier networks propagate to later ones, and an information bottleneck emerges as later stages operate on increasingly abstract representations. Our goal is to learn a single, reusable sensor-level backbone on raw 3D charge and attach a compact set-prediction head that also operates on point-level features, thereby removing detector-tuned heuristics, large label requirements, and simulation bias.

%% file: secs/3_method.tex
\section{Method}

\subsection{Encoder}

Panda is built around a shared point-native hierarchical encoder, Point Transformer V3 \cite{wu2024pointtransformerv3simpler} and a set of lightweight task heads which operate on raw voxels from LArTPC images. The same encoder is used for self-distilled pre-training, dense semantic segmentation, and instance-level reconstruction.

\subsubsection{Input representation}
Each event is represented as a set of voxelized 3D points $\{(x_i, y_i, z_i, q_i)\}_{i=1}^N$ corresponding to charge depositions in the LArTPC volume on a 3~mm grid, where $(x_i,y_i,z_i)$ denotes the 3D position and $q_i$ is the energy, in MeV, associated with that point.

\begin{figure}[t]
    \centering
    \includegraphics[width=0.95\linewidth]{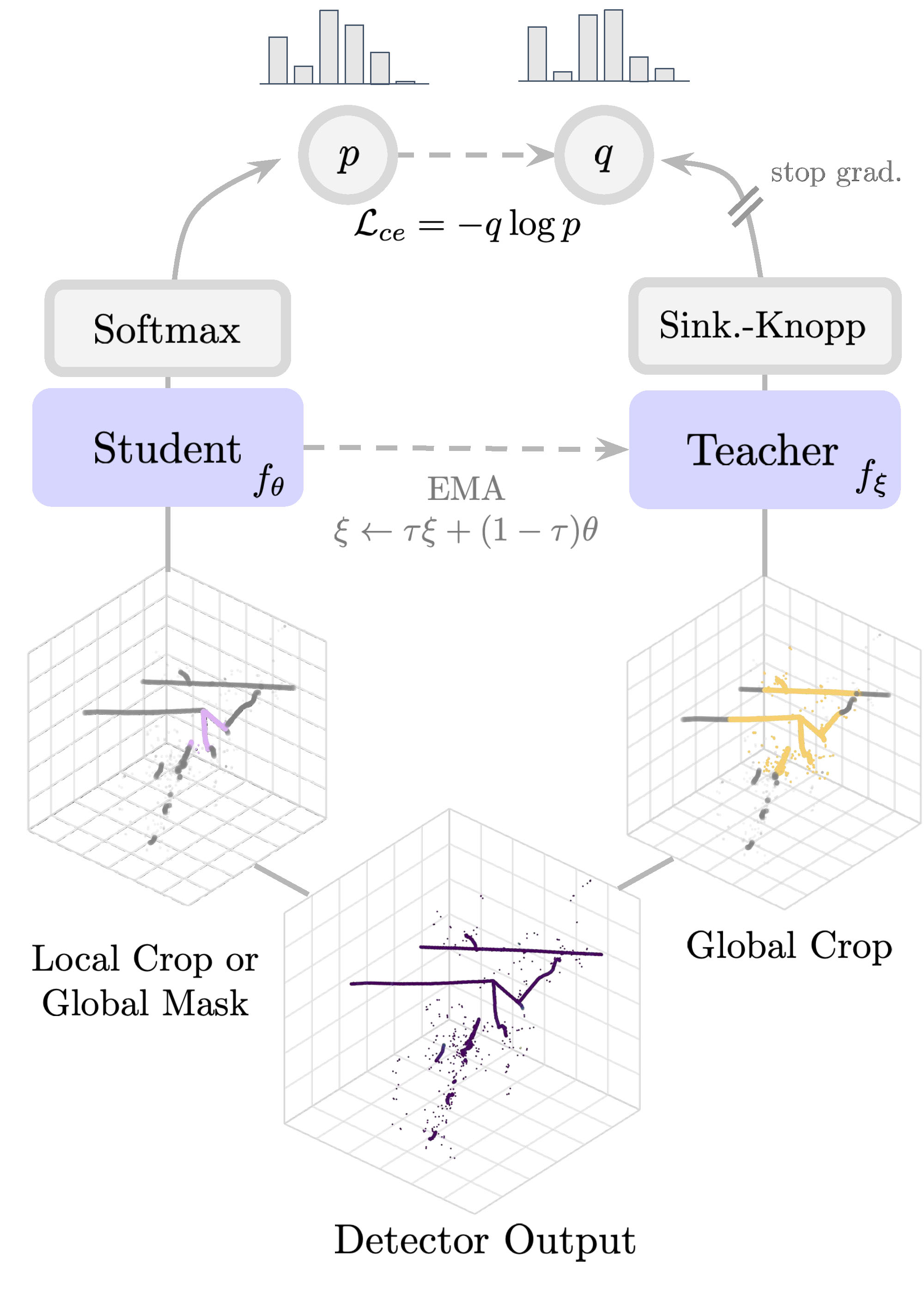}
    \caption{\textbf{Self-distillation setup.}
    Global and local/masked views are fed to separate student ($f_\theta$) and teacher ($f_\xi$) encoders that share the Panda backbone. Each network produces a per-point distribution over prototypes. Cross-entropy losses align
    student predictions on local and masked views with the corresponding
    unmasked global teacher predictions, enforcing consistent prototype
    assignments across views. The teacher's weights is updated as an exponential moving average of the student's weights, with $\tau$ scheduled from 0.994 to 1.0 over the length of pretraining.}
    \label{fig:selfdist_setup}
\end{figure}

\subsubsection{Point-native hierarchical encoder}
\label{sec:encoder}
The backbone is a five-stage U-Net–style hierarchy built from sparse 3D convolutions~\cite{choy20194dspatiotemporalconvnetsminkowski} and self-attention over local point sets, following Point Transformer V3~\cite{wu2024pointtransformerv3simpler}. It mirrors the Sparse UResNet used for semantic segmentation in LArTPC experiments~\cite{drielsma2021scalableendtoenddeeplearningbaseddata}, with the key addition of self-attention after each sparse convolution.

We initialize a sparse grid at the native LArTPC voxel size. At each stage, 3$\!\times\!$3$\!\times\!$3 sparse convolution blocks aggregate local geometry, then points are serialized with alternating space-filling curves (Hilbert/Morton and transverse) into length 256 sequences for multi-head self-attention~\cite{vaswani2017attention}, yielding localized receptive fields. Between stages, grid pooling with stride of 2 aggregates features within 2$\!\times\!$2$\!\times\!$2 cells to form coarser grids, followed by PTv3 blocks. We choose a 91M parameter encoder with 5 scales (corresponding to 3 mm, 6 mm, 12 mm, 24 mm and 48 mm), block depths $(3, 3, 3, 9, 3)$, channel widths $(48, 96, 192, 384, 512)$, and a fixed serialized patch size of 256 at all levels. 

A decoder is used only as a lightweight adapter for semantic segmentation; otherwise we upcast and concatenate features from the top $L$ encoder levels. The resulting multi-scale features $\{F^{(l)}\}_{l=0}^{L}$ serve as the shared representation for pre-training and all downstream tasks.

Between each stage of the UNet, we perform grid pooling with a stride of 2 voxels in each spatial dimension, with the initial grid size corresponding to a single voxel. Features within each 2$\times$2$\times$2 block are aggregated and projected through a small MLP to define the coarser sparse grid for the next stage. After grid pooling, a number of PTv3 blocks is applied. 

\subsection{Decoder}

We optionally include a 16M parameter UNet-like multi-scale decoder. This decoder progressively upsamples encoded features at each depth, combined with corresponding skip-connections from the encoder, and results in point-level embeddings. It mirrors the encoder path, with the exceptions of each depth containing just two PTv3 blocks, and the final embedding resolution being 64 instead of the encoder's intial 48.

\subsubsection{Semantic segmentation}
For dense semantic segmentation, we apply a linear classification head on per-point features outputted from either the encoder-only backbone via upcasting and concatenation, or the multi-scale decoder.

\subsubsection{Panoptic segmentation}

The self-distilled encoder is trained to cluster recurring local motifs in feature space. This objective encourages invariance across location and context, so two particles of the same type on opposite sides of the detector can have nearly identical embeddings. For physics reconstruction, we additionally need to (i) disentangle individual particles that overlap in space and in latent space, and (ii) group causally related particles into interaction-level objects. We therefore add a lightweight set-prediction head on top of the pretrained backbone.

We cast particle and interaction reconstruction as a set prediction problem, and base our model off of Mask2Former \cite{m2f} and OneFormer3D~\cite{kolodiazhnyi2023oneformer3dtransformerunifiedpoint}. 

\paragraph{Thing--stuff separation.}
A hallmark of LArTPC data is the presence of ubiquitous low-energy depositions (LEDs): small, spatially scattered blobs of energy deposition that are not individually countable objects. These can correlate with individual particles but are not generally useful as distinct instances. Let $E \in \mathbb{R}^{N \times D}$ ($D=256$) be the up-casted per-point embeddings after dimensionality reduction via a two-layer MLP. Before introduction to the set-prediction head, we perform binary classification on each point as either ``LED" (stuff) versus candidate objects (things) using a small MLP on top of $E$. Unlike in tradition panoptic segmentation, where there can be multiple ``thing'' classes, LEDs are our sole ``thing'' class, and so a single MLP is satisfactory. Points predicted as LEDs are excluded from the instance head and treated as background / stuff.

\paragraph{Mask prediction and query specialization.}

Queries act as instance prototypes and are refined across three decoder blocks. At each block, we bias cross-attention so that each query focuses on the points it is already likely to own, while still allowing competition and correction.

Let $\{q^{(l)}_k\}_{k=1}^Q$ be the query embeddings at decoder block $l$ and again $E \in \mathbb{R}^{N \times D}$ the encoded features after dimensionality reduction. For each query $k$, a small MLP maps $q^{(l)}_k$ to a mask embedding $m^{(l)}_k \in \mathbb{R}^D$. We compute point--query affinities 
$z^{(l)}_{k,i} = \langle m^{(l)}_k,\, E_i \rangle,$
and corresponding soft assignment scores
$
\hat{p}^{(l)}_{k,i} = \sigma\!(z^{(l)}_{k,i}),
$
interpreted as the current belief that point $i$ belongs to query $k$ at block $l$.

Taking inspiration from \cite{m2f,park2025fm4npp}, we convert these scores into an additive attention bias
\[
b^{(l)}_{k,i} = \log\!\left(\hat{p}^{(l)}_{k,i} + \varepsilon\right), \quad \varepsilon = 10^{-6},
\]
detach $\hat{p}^{(l)}_{k,i}$ from the gradient, and add $b^{(l)}_{k,i}$ to the pre-softmax logits used for the cross-attention step that updates $q^{(l)}_k$ from $E$. Cross-attention is followed by self-attention among queries to promote query specialization, and finally a feed-forward network that operates on each query independently. This iterative procedure, which is repeated three times, encourages each query to converge to a coherent, possibly non-local subset of points, while retaining flexibility to reassign points across queries.

The queries themselves are learnable and are associated with learnable positional embeddings; both are initialized randomly. We use 32 learned queries for particle segmentation and 12 for interaction segmentation. At each layer of the decoder, these positional embeddings are added onto the query embeddings. To aid in disentangling particle instances from one another, we add a learned positional encoding to each point feature before cross attention.

At the final block $L$, the scores $\pi_{k,i} \equiv \hat{p}^{(L)}_{k,i}$ serve as the predicted soft mask for query $k$ over points $i=1,\dots,N$.

\paragraph{Training objective.}
We supervise the instance head with a matching-based loss over predicted queries and ground-truth instances.

Let $\mathcal{G}$ be the set of ground-truth instances (particles or interaction groups). Each $g \in \mathcal{G}$ is defined by:
(i) a binary mask $y_{g,i} \in \{0,1\}$ over points, and
(ii) a class label $c_g$ from a set of particle classes.
Each query $k \in \{1,\dots,Q\}$ produces:
(i) class probabilities $p_k$ over the same label set plus a no-object label $\varnothing$, and
(ii) a soft mask $\pi_{k,i} \in [0,1]$.

For every pair $(k,g)$, we define a matching cost
\[
\mathcal{C}(k,g)
= \lambda_{\text{mask}}\,\mathcal{L}_{\text{mask}}(\pi_k, y_g)
+ \lambda_{\text{cls}}\,\mathcal{L}_{\text{cls}}(p_k, c_g),
\]
where $\mathcal{L}_{\text{cls}}$ is cross-entropy on $c_g$ (excluding $\varnothing$),
and $\mathcal{L}_{\text{mask}}$ combines focal \cite{lin2018focallossdenseobject} and Dice \cite{Sudre_2017} losses:
\[
\mathcal{L}_{\text{mask}}(\pi_k, y_g)
= \lambda_{\text{focal}}\,\mathcal{L}_{\text{focal}}(\pi_k, y_g)
+ \lambda_{\text{dice}}\,\mathcal{L}_{\text{dice}}(\pi_k, y_g).
\]
We use $\lambda_{\text{cls}}=\lambda_{\text{mask}}=\lambda_{\text{focal}}=\lambda_{\text{dice}}=1.0.$ As in DETR~\cite{detr}, the Hungarian algorithm is used to find an optimal one-to-one assignment $\phi: \mathcal{G} \to \{1,\dots,Q\}$
minimizing the total cost. Queries not in the image of $\phi$ are treated as unmatched. For each ground-truth instance $g$ matched to query $k = \phi(g)$, we apply both classification and mask losses:
\[
\mathcal{L}_{\text{matched}}
= \sum_{g \in \mathcal{G}} \Big[
\lambda_{\text{cls}}\,\mathcal{L}_{\text{cls}}(p_{\phi(g)}, c_g)
+ \lambda_{\text{mask}}\mathcal{L}_{\text{mask}}(\pi_{\phi(g)}, y_g)
\Big],
\]
now with $\lambda_{\text{cls}}=2.0$, $\lambda_{\text{mask}}=1.0$, $\lambda_{\text{dice}}=5.0$, and $\lambda_{\text{focal}}=2.0$
For unmatched queries $\mathcal{U} = \{1,\dots,Q\} \setminus \phi(\mathcal{G})$, we supervise the no-object label with a down-weighted classification loss:
\[
\mathcal{L}_{\text{unmatched}}
= \lambda_{\varnothing} \sum_{k \in \mathcal{U}}
\mathcal{L}_{\text{cls}}(p_k, \varnothing),
\]
with $\lambda_{\varnothing} = 0.5$. The total instance loss at the final decoder block is thus
\[
\mathcal{L}_{\text{inst}}^{(L)} =
\mathcal{L}_{\text{matched}} + \mathcal{L}_{\text{unmatched}}.
\]
We perform deep supervision~\cite{lee2015deeply}, applying the same matching and loss at each decoder block $l$, using that block’s masks $\hat{p}^{(l)}_{k,i}$ and class predictions. The final training objective is the sum
\[
\mathcal{L}_{\text{inst}} =
\sum_{l=1}^{L}
\left(
\mathcal{L}_{\text{matched}}^{(l)} + \mathcal{L}_{\text{unmatched}}^{(l)}
\right),
\]
where $\mathcal{L}_{\text{matched}}^{(l)}$ ($\mathcal{L}_{\text{unmatched}}^{(l)}$) is the matched (unmatched) loss computed at the $l^\text{th}$ layer of the set prediction module.

\paragraph{Inference.}
At inference, the $k$th query outputs:
(i) class probabilities $p_k \in \Delta^{C+1}$ over $C$ particle classes plus a no-object label $\varnothing$, and
(ii) per-point mask logits $m_{k,i} \in \mathbb{R}$ for points $i = 1,\dots,N$. We first assign classes to each query $c_k=\arg\max p_k(c),$ discarding those classified as non-objects. For the remaining queries, we form soft masks $\pi_{k,i} = \sigma(m_{k,i}),$ and define their point indices $\Omega_k = \{ i \mid \pi_{k,i} \ge \tau_m \}$ with $\tau_m=0.5.$ Queries with $\lvert \Omega_k \rvert < K_{\min}=2$, containing too few points to represent meaningful objects, are removed. For each surviving query, we define a confidence score
\[
s_k
= p_k(c_k) \cdot
\frac{1}{\lvert \Omega_k\rvert}
\sum_{i \in \Omega_k} \pi_{k,i},
\]
which is the product of class confidence and the average confidence in $\Omega_k.$
Within each class $c$, we perform mask-based non-maximum suppression using matrix NMS \cite{wang2020solov2dynamicfastinstance}. This yields a consistent set of particle-level and interaction-level instances from a single pretrained backbone and compact instance head.

\subsection{Pre-training objective}
\label{sec:selfdistill}

\begin{figure*}[t]
    \centering
    \includegraphics[width=\linewidth]{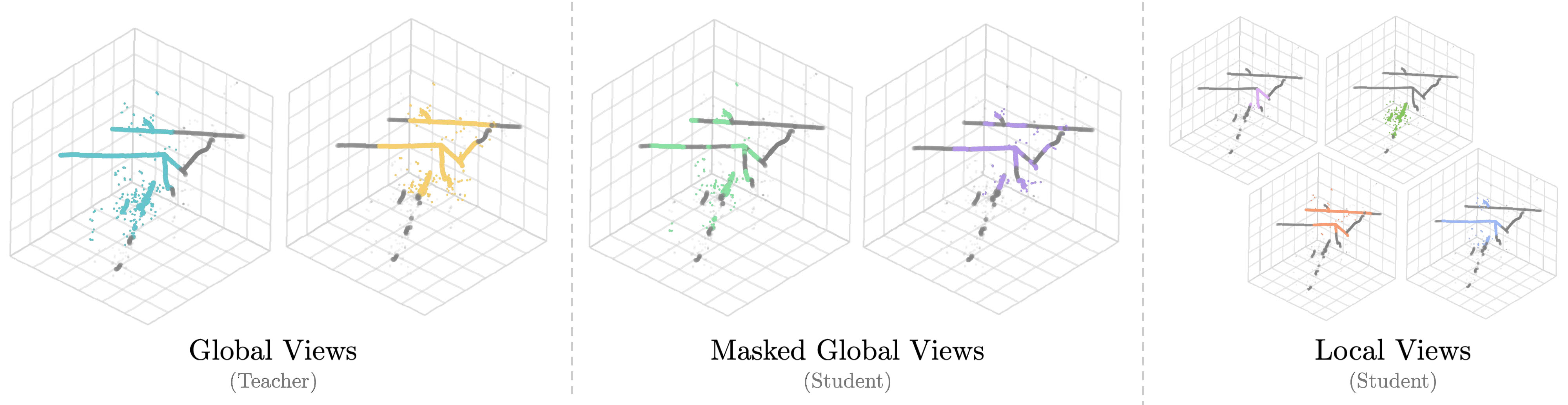}
    \caption{\textbf{Global, local and masked views used in Panda.}
    Example LArTPC event with two global crops and several local views.
    Local crops cover contiguous regions of the event. Masked global
    views hide large subsets of patches from the student while the
    teacher sees the full event.}
    \label{fig:selfdist_views}
\end{figure*}

We use a decoder-free, prototype-based self-distillation scheme to pre-train Panda. This is a discriminative self-supervised strategy (illustrated in Fig.~\ref{fig:selfdist_setup}) inspired by DINO and iBOT~\cite{caron2021emerging,zhou2022ibotimagebertpretraining,oquab2024dinov2learningrobustvisual} and Sonata~\cite{wu2025sonataselfsupervisedlearningreliable}, whereby prototype distributions over image patches are learned in a teacher-student setup. Notably, we only operate on point-level outputs from the model backbone, without any per-image \texttt{[cls]} tokens, i.e. global summary tokens used in Vision Transformers. We eschew \texttt{[cls]} tokens because, unlike natural images, our LArTPC images contain causally unrelated particle interactions with their own local topologies. 

A student encoder $f_\theta$ and an EMA teacher encoder $f_\xi$ share the Panda backbone. Following encoding by the backbone network, a MLP projects each upcasted per-point feature to a low dimensional latent space, and a shared prototype layer maps these features to prototype distributions that define the self-distillation targets.

\subsubsection{Prototypes as clusters on a hypersphere}
At the core of our objective is a geometric similarity search performed on a set of $K$ learnable prototypes. Following encoding by the backbone network, a MLP projects each upcasted per-point feature to a low dimensional latent space ($D=256$). Let $z \in \mathbb{R}^D$ be this feature embedding output. We introduce a weight matrix $W \in \mathbb{R}^{K \times D}$ ($K=4096$), where each row vector $w_k$ represents a prototype (i.e., a cluster center) in the feature space.

Unlike a standard linear layer where the output score is $s_k = w_k \cdot z$, we enforce a normalization constraint to project the optimization onto a unit hypersphere. We apply $L_2$ normalization to the input features, yielding $\hat{z} = z / \|z\|_2$, and use Weight Normalization~\cite{salimans2016weightnormalizationsimplereparameterization} on $W$ to fix the $L_2$ norm of each prototype vector to $1$ (i.e., $\|w_k\|_2 = 1$). Consequently, the dot product becomes a measure of cosine similarity:
\begin{equation*}
    s_k = \hat{z} \cdot w_k = \cos(\theta_k),
\end{equation*}
with $\theta_k$ being the angle between $\hat{z}$ and $w_k$.
During training, the network simultaneously updates the encoder to produce features $\hat{z}$ that align with specific prototypes, and updates the prototypes $w_k$ to better represent groups of semantically similar features.

The student encoder $f_\theta$ and EMA teacher encoder $f_\xi$ share the Panda backbone and produce per-point embeddings at $12$\,mm (4 voxel) resolution created by upcasting and concatenating as in Sec.~\ref{sec:encoder} until the third scale, followed by the shared prototype layer with $K=4096$ learnable prototypes.

\subsubsection{Global, local, and masked views}
For each unlabeled event, we sample two global crops containing $40$--$100\%$ of the total number of voxels and several smaller local crops with $10$--$40\%$, as illustrated in Fig.~\ref{fig:selfdist_views}. All crops are augmented only by rotations, flips, centering, and a mild Gaussian coordinate jitter ($\sigma=0.5$ grid size units, roughly half a voxel). We also apply a small multiplicative jitter to charge: $q_i \rightarrow q_i \cdot u$ with $u \sim \mathcal{N}(1,\sigma=0.05)$. 

Following Sonata~\cite{wu2025sonataselfsupervisedlearningreliable}, we use two types of local views. The first is a contiguous crop: we sample a random anchor point in the image and take nearby points until we reach a target fraction of the event ($10$--$40\%$ of the total points). For each local view, the student must predict prototype distributions that match those of the corresponding region in an unmasked global teacher view. Teacher predictions are computed only on global crops and student predictions on local crops. This defines the local cross-entropy term $\mathcal{L}_\mathrm{local}$, which aligns prototype distributions between local views and the corresponding regions of the unmasked global teacher view.

The second local view is a masked global crop. We start from a global view, partition it into voxelized grid patches, and then drop a large fraction of the patches on the student side, replacing them with a learned ``masked point'' embedding and an additional coordinate jitter on the masked points. A masked variant of the global views hides patches from the student while keeping them for the teacher. Masked patch size increases from $\sim 7$ voxels ($2.1$ cm) to $\sim 50$ voxels ($15$ cm) and the masked ratio from $50\%$ to $90\%$ over the first $5\%$ of training, with a slightly stronger ($\sim$ 1 voxel) additive jitter on the masked points to reduce the possibility of the model naively clustering based on position. The teacher still sees the full unmasked global crop. A mask loss (also cross-entropy) $\mathcal{L}_\mathrm{mask}$ operates on these masked global views. Here, only masked points contribute, and the student is required to match the unmasked global teacher. This mask loss is computed for both global crops. These choices preserve physically meaningful geometry while forcing the model to rely on stable local charge patterns.

Let $q_j$ denote the teacher’s prototype distribution for a point $j$ in a global view (with temperature $\tau_t$ linearly increased from $0.04$ to $0.07$), and $p_i$ the student’s distribution for the corresponding point in an augmented view (with temperature $\tau_s = 0.1$). The overall self-distillation objective is
\[
\mathcal{L}_\mathrm{SSL}
= \lambda_\mathrm{local}\,\mathcal{L}_\mathrm{local}
+ \lambda_\mathrm{mask}\,\mathcal{L}_\mathrm{mask},
\]
with $\lambda_\mathrm{local}=\lambda_\mathrm{mask}=0.5$.
This setup enforces consistent prototype assignments across several scales, occlusions, and views, and produces the sensor-level representations used by our semantic and instance reconstruction heads.

\subsubsection{Optimization and collapse prevention}
The optimization objective (visualized in Fig.~\ref{fig:selfdist_setup}) is to minimize the cross-entropy loss $\mathcal{L}_{ce}$ between the teacher's output distribution $q$ and the student's output distribution $p$. This loss can be decomposed into the entropy of the teacher $h(q)$ and the KL-divergence between teacher and student (ignoring the summing over all elements):
\begin{align*}
    \mathcal{L}_{ce}(q, p) &= -q\log p\\
    &= \underbrace{-q\log q}_{h(q)} + \underbrace{q\log q - q\log p}_{D_{KL}(q\parallel p)}\\
    &= h(q) + D_{KL}(q\parallel p).
\end{align*}

Without any constraints, self-distillation is prone to \textit{representation collapse}, where the loss becomes constant regardless of the input. This typically manifests in two modes: \textbf{uniform collapse}, where the model outputs a uniform distribution for all inputs ($h(q) = \log K$); and \textbf{single-prototype collapse}, where the model maps all inputs to a single prototype ($h(q) = 0$).
In both cases, $D_{KL}(q\parallel p)=0$, resulting in vanishing gradients due to the stop gradient applied to the teacher. To prevent this, we follow DINO~\cite{caron2021emerging,oquab2024dinov2learningrobustvisual} in employing asymmetric sharpening and centering. For the student, we sharpen the distribution using a low temperature $\tau_s = 0.1$. For the teacher, we apply a dynamic temperature schedule warming up from $0.04$ to $0.07$ over the first $5\%$ of training. Furthermore, we replace the teacher's regular Softmax operation with the Sinkhorn-Knopp (SK) batch normalization from SwAV~\cite{swav}, a centering-like operation that ensures that the teacher utilizes all prototypes equally within the batch while maintaining the semantic structure of the clusters. Together, these two constraints push the prototype distributions away from uniform and single-prototype collapse.

To provide a stable target for the student, the teacher's weights $\xi$ are instead updated as a moving average of the student's weights $\theta$; that is, at each iteration $\xi \leftarrow \tau \xi + (1-\tau)\theta$. We schedule $\tau$ from $0.994$ to $1.0$ over the length of training.

\subsection{Data} We pre-train on PILArNet-M~\cite{polarmae}, a dataset of 1.2M LArTPC events. We procured pixel-level particle information for all events, information that was not originally included in the dataset release, allowing us to perform particle-level panoptic segmentation. For all images, we perform the same data transformations as in PoLAr-MAE~\cite{polarmae}, most notably removing all points $<0.13~$MeV. This value, which corresponds to 20\% of the lowest expected signal value, mimics realistic readout hardware performance and reduces overall training time.

\begin{table*}[t]
    \begin{minipage}{0.6\textwidth}
    \centering
        \tablestyle{1.9pt}{1.08}
        \input{table/seg_data_efficiency}
        \vspace{-3mm}
        \captionof{table}{\textbf{Data efficiency for semantic segmentation.}}
        \label{tab:semseg_dataeff}
        \vspace{1mm}
\end{minipage}
\hspace{0mm}
\begin{minipage}{0.40\textwidth}
    \centering
    \vspace{-3mm}
    \resizebox{1.\textwidth}{!}{
    \input{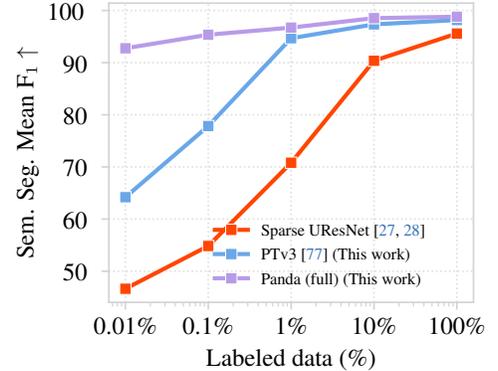}}
    \vspace{-9mm}
    \captionof{figure}{\textbf{Data efficiency curve for semantic segmentation.}}
    \label{fig:semseg_scaling}
    \vspace{1mm}
    \end{minipage}
\end{table*}

\begin{table}[t]
    \centering
    \vspace{-3mm}
    \resizebox{\linewidth}{!}{
        \tablestyle{0.0pt}{1}
        \hspace{0mm}
        \input{table/seg_high_stats}
        }
        \vspace{-3mm}
        \captionof{table}{\textbf{Parameter efficiency for semantic segmentation.} We fine-tune Panda pre-trained on 100\% of the dataset (1M images).}
        \label{tab:semseg_high_stats}
        \vspace{-3mm}
\end{table}

\subsection{Training} 
We pretrain Panda for $10$M event-samples (equivalent to $10$ passes over a $1$M-event dataset), corresponding to $208{,}333$ optimizer steps at an effective batch size of $48$ on 4$\times$A100 40 GB GPUs. Fine-tuning uses $20$M event-samples ($416{,}666$ steps at batch $48$) for both semantic and instance segmentation, using 4$\times$A100 for the former and 8$\times$A100 for the latter. We optimize with AdamW~\cite{loshchilov2019decoupledweightdecayregularization} with $(\beta_1,\beta_2)=(0.9,0.999)$, a base learning rate of $2.6\times10^{-3}$ with $5\%$ linear warmup followed by cosine decay to $10^{-3}$ of the base rate, and cosine-scheduled weight decay that increases from $0.04$ to $0.20$ during pre-training and is fixed at $0.01$ for fine-tuning; bias and normalization parameters are excluded from decay. We apply depth-wise learning-rate decay with factor $\lambda=0.9$ for pre-training and instance-level fine-tuning and $\lambda=0.97$ for semantic-segmentation fine-tuning, stochastic depth~\cite{huang2016deepnetworksstochasticdepth} with rate $0.3$, and per-block layer scaling~\cite{touvron2021goingdeeperimagetransformers} for attention initialized at $10^{-5}$. EMA teacher parameters use momentum increasing from $0.994$ to $1.0$, with student temperature $\tau_s=0.1$ and teacher temperature $\tau_t$ rising from $0.04$ to $0.07$. During pre-training, we periodically train a linear probe on frozen features for semantic segmentation and select the final checkpoint either as the last epoch (if validation $F_1$ remains stable) or the best linear-probe $F_1$ (if representational collapse is detected for very small datasets $K\in\{10^2,\dots,10^4\}$ that are heavily oversampled). Fine-tuning reuses only random rotations and flips as augmentations.

\subsection{Evaluation} 
\label{sec:eval}

We first examine representation geometry with t-SNE~\cite{JMLR:v9:vandermaaten08a} on unit-normalized point embeddings from 1,000 images, visualizing clusters when colored by both particle type and geometric motif. For dense prediction, we perform semantic segmentation exclusively on these motifs and report macro $F_1$ and per-class scores. To study data efficiency and scaling, we run two sweeps: \emph{PT$_{100\%}$+FT$_{K\%}$}, where a backbone pretrained on 100\% of the dataset (1M events) is fine-tuned on $K\!\in\!\{0.01\%,0.1\%,1\%,10\%,100\%\}$ of the full dataset, and \emph{PT$_{K\%}$+FT$_{K\%}$}, where both pretraining and fine-tuning use the same $K$. Parameter efficiency is probed with three modes for adaptation: linear probing, decoder-only fine-tuning with frozen features, and full fine-tuning. We additionally train from scratch with the same architecture and schedule to disentangle architectural improvements from those attributed to pretraining.

For instance-level tasks, we evaluate particle- and interaction- level panoptic segmentation~(Fig. \ref{fig:teaser}, bottom center/right). We fine-tune a small (4M parameter) Mask2Former~\cite{m2f}-style set-prediction head on top of the frozen PT$_{100\%}$ backbone across $K\!\in\!\{0.01\%,\dots,100\%\}$ to probe data efficiency, and perform full fine-tuning and from-scratch training at $K\!=\!100\%$ to examine parameter efficiency. We evaluate via Panoptic Quality (PQ)~\cite{kirillov2019panopticsegmentation} and Adjusted Rand Index (ARI)~\cite{Hubert1985}; PQ, prevalent in the panoptic segmentation community, evaluates both the recognition quality (RQ) and segmentation quality (SQ) of objects in panoptic segmentation, while the ARI, ubiquitous in LArTPC reconstruction literature, measures the similarity between predicted and true clusterings based on pairwise label consistency.

Across all experiments, we report results under matched training budgets to understand improvements for data-efficiency and task transfer.

%% file: table/seg_data_efficiency.tex
\begin{tabular}{lccccc@{\hspace{0.5\tabcolsep}}c@{\hspace{0.5\tabcolsep}}ccccc}
\toprule
Semantic Segmentation  & \multicolumn{5}{c}{PT$_{K\%}$ + FT$_{K\%}$} &  & \multicolumn{5}{c}{PT$_\text{100\%}$ + FT$_{K\%}$} \\
  \cmidrule(lr){1-1} \cmidrule(lr){2-6} \cmidrule(lr){7-12}
Method / $K\%$ & 0.01\% & 0.1\% & 1\% & 10\% & 100\% &  & 0.01\% & 0.1\% & 1\% & 10\% & 100\% \\
\midrule
\multicolumn{12}{l}{\textit{Supervised}} \\
\sota~UResNet~\cite{choy20194dspatiotemporalconvnetsminkowski,domine2020scalable} & 46.6 & 54.8 & 70.8 & 90.4 & \cellcolor[HTML]{F4F4FF}95.6 & \footnotesize{=} & 46.6 & 54.8 & 70.8 & 90.4 & \cellcolor[HTML]{F4F4FF}95.6 \\
\thiswork~PTv3~\cite{wu2024pointtransformerv3simpler}  & 64.2 & 77.8 & \cellcolor[HTML]{F4F4FF}94.7 & \cellcolor[HTML]{E3E3FF}97.3 & \cellcolor[HTML]{E3E3FF}98.2 & \footnotesize{=} & 64.2 & 77.8 & \cellcolor[HTML]{F4F4FF}94.7 & \cellcolor[HTML]{F4F4FF}97.3 & \cellcolor[HTML]{E3E3FF}98.2 \\
\midrule
\multicolumn{12}{l}{\textit{Self-supervised}} \\
\otherwork~PoLAr-MAE~\cite{polarmae}~(dec.) & 71.3 & 80.6 & 84.6 & 83.4 & 83.4 &  & -- & -- & -- & -- & 83.4 \\
\otherwork~PoLAr-MAE~\cite{polarmae}~(full) & 73.5 & 84.0 & 86.6 & 86.5 & 85.7 &  & -- & -- & -- & -- & 85.7 \\
\thiswork~Panda (lin.) & \cellcolor[HTML]{F4F4FF}79.1 & \cellcolor[HTML]{F4F4FF}90.1 & 93.2 & \cellcolor[HTML]{F4F4FF}93.7 & 93.9 &  & \cellcolor[HTML]{F4F4FF}92.2 & \cellcolor[HTML]{F4F4FF}93.6 & 93.8 & 93.9 & 93.9 \\
\thiswork~Panda (dec.) & \cellcolor[HTML]{E3E3FF}83.5 & \cellcolor[HTML]{E3E3FF}92.8 & \cellcolor[HTML]{E3E3FF}95.9 & \cellcolor[HTML]{E3E3FF}97.3 & \cellcolor[HTML]{E3E3FF}98.2 &  & \cellcolor[HTML]{E3E3FF}92.6 & \cellcolor[HTML]{E3E3FF}95.2 & \cellcolor[HTML]{E3E3FF}96.2 & \cellcolor[HTML]{E3E3FF}97.8 & \cellcolor[HTML]{E3E3FF}98.2 \\
\thiswork~Panda (full) & \cellcolor[HTML]{D6D6FF}\textbf{85.2} & \cellcolor[HTML]{D6D6FF}\textbf{93.7} & \cellcolor[HTML]{D6D6FF}\textbf{96.4} & \cellcolor[HTML]{D6D6FF}\textbf{97.7} & \cellcolor[HTML]{D6D6FF}\textbf{98.8} &  & \cellcolor[HTML]{D6D6FF}\textbf{92.8} & \cellcolor[HTML]{D6D6FF}\textbf{95.4} & \cellcolor[HTML]{D6D6FF}\textbf{96.7} & \cellcolor[HTML]{D6D6FF}\textbf{98.5} & \cellcolor[HTML]{D6D6FF}\textbf{98.8} \\
\bottomrule
\multicolumn{12}{r}{\sota~Prev. SOTA\hspace{2\tabcolsep}\thiswork~This work}  \\
\end{tabular}

%% file: table/seg_high_stats.tex
\hspace{-5.7mm}
\begin{tabular}{lrc@{\hspace{0.5\tabcolsep}}c@{\hspace{0.5\tabcolsep}}ccccc}
\toprule
Semantic Segmentation & \multicolumn{1}{c}{Param.}  & \multicolumn{7}{c}{$F_1$} \\
\cmidrule(lr){1-1} \cmidrule(lr){2-2} \cmidrule(lr){3-9} 
Method & & Macro & ~ & Shower & ~Track~ & Michel & ~Delta~ & ~LED~ \\
\midrule
\multicolumn{9}{l}{\textit{Supervised}} \\
\sota~UResNet~\cite{choy20194dspatiotemporalconvnetsminkowski,domine2020scalable}  & 11.0~M & \cellcolor[HTML]{F4F4FF}95.6 &&  \cellcolor[HTML]{E3E3FF}99.1 & \cellcolor[HTML]{F4F4FF}99.5 &  91.3 & 90.8 & \cellcolor[HTML]{D6D6FF}\textbf{97.3} \\
\thiswork~PTv3~\cite{wu2024pointtransformerv3simpler} & 107.1~M & \cellcolor[HTML]{E3E3FF}98.2 & & \cellcolor[HTML]{D6D6FF}\textbf{99.4} & \cellcolor[HTML]{E3E3FF}99.8 & \cellcolor[HTML]{F4F4FF}97.6 & \cellcolor[HTML]{E3E3FF}97.0 & \cellcolor[HTML]{F4F4FF}97.0 \\
\midrule
\multicolumn{9}{l}{\textit{Self-supervised}} \\
\otherwork~PoLAr-MAE~\cite{polarmae}~(dec.) & 4.4~M & 83.4 & & 96.8 & 97.3 & 73.9 & 62.8 & 86.2 \\
\otherwork~PoLAr-MAE~\cite{polarmae}~(full) & 24.4 M & 85.7 & & 97.3 & 97.4 & 83.2 & 64.0 & 86.7 \\
\thiswork~Panda (lin.) & 0.01~M & 93.9 & & \cellcolor[HTML]{F4F4FF}98.7 & 99.1 & 95.3 & 82.0 & 94.5 \\
\thiswork~Panda (dec.) & 16.3~M & \cellcolor[HTML]{E3E3FF}98.2 & & \cellcolor[HTML]{D6D6FF}\textbf{99.4} & \cellcolor[HTML]{E3E3FF}99.8 & \cellcolor[HTML]{E3E3FF}98.0 & \cellcolor[HTML]{F4F4FF}96.6 & 96.9 \\
\thiswork~Panda (full) & 107.1~M & \cellcolor[HTML]{D6D6FF}\textbf{98.4} & & \cellcolor[HTML]{D6D6FF}\textbf{99.4} & \cellcolor[HTML]{D6D6FF}\textbf{99.9} & \cellcolor[HTML]{D6D6FF}\textbf{98.5} & \cellcolor[HTML]{D6D6FF}\textbf{97.4} & \cellcolor[HTML]{E3E3FF}97.1 \\
\bottomrule
\multicolumn{9}{r}{\sota~Prev. SOTA\hspace{2mm}\thiswork~This work}  \\
\end{tabular}

%% file: secs/4_results.tex
\begin{figure*}[th!]
    \centering
    \includegraphics[width=1\linewidth]{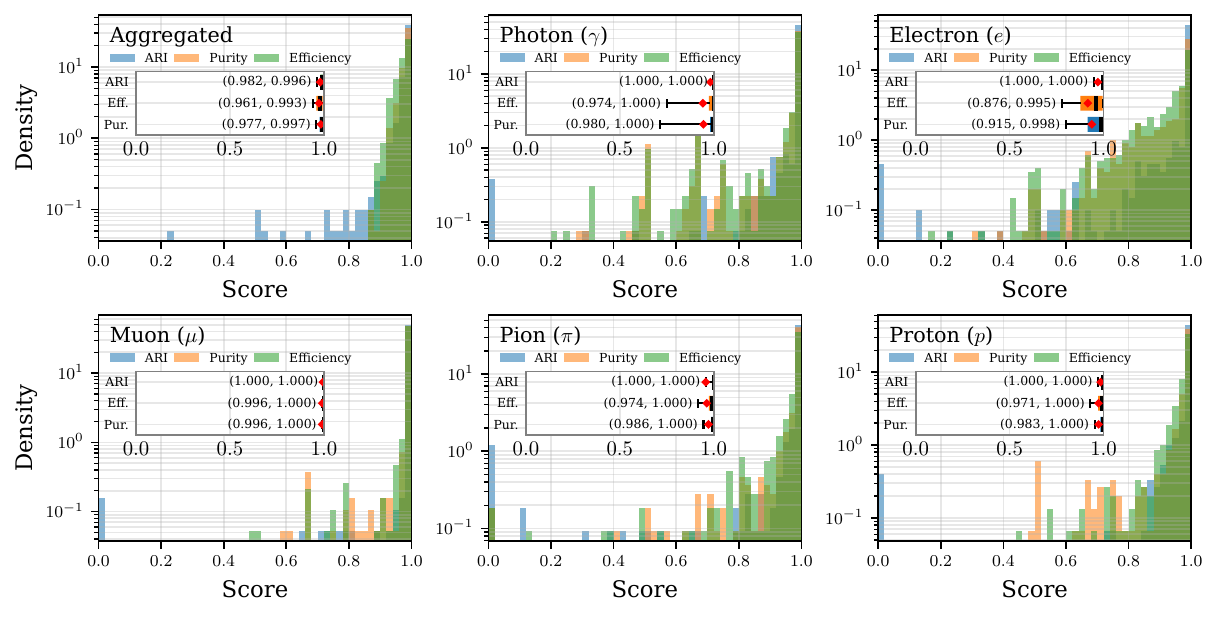}
    \vspace{-8mm}
    \caption{\textbf{Per-particle panoptic segmentation metrics.} We provide overall ARI/purity/efficiency (``aggregated``), as well as particle-specific metrics. Histograms are plotted on a log-scale. These statistics are summarized with box plots; here, the red diamonds correspond to the mean, the black line the median, the boxes the interquartile range (IQR), and the whiskers the 10th and 90th percentiles. The IQR is printed numerically in parenthesis.}
    \label{fig:ari_summary_boxplot}
\end{figure*}
\section{Results}
\label{sec:results}
We evaluate Panda under the protocols in Sec.~\ref{sec:eval}, focusing on (i) the structure of the learned representation, (ii) data and parameter efficiency for dense semantic labeling, and (iii) particle– and interaction–level panoptic segmentation.

\paragraph{Feature probing with t-SNE.}
In Fig.~\ref{fig:tsne}, we look at a t-SNE plot generated from point embeddings from 1,000 images and see that the inter-class diversity and intra-class multi-modality introduced in Sec.~\ref{sec:intro} are very well-separated. We observe that not only are individual particles well-separated, but so too are the same particle created from different physics processes. We also notice substantial overlap within the t-SNE between electrons ($e$) and photons ($\gamma$), as well as with muons ($\mu$) and pions ($\pi$). This reflects well-known ambiguities within LArTPC events; $\gamma$-initiated electromagnetic (EM) showers without a resolvable conversion gap and reliable energy deposition (dE/dx) patterns over the first few centimeters of the trajectory are effectively indistinguishable from a single electron~\cite{Acciarri_2014,Adams_2020}, and long tracks that are non-interacting or not fully contained within the detector can make $\mu/\pi$ separation challenging~\cite{muon2021}.

\paragraph{Semantic segmentation data efficiency.}
Table~\ref{tab:semseg_dataeff} reports macro $F_1$ across labeled fractions for the semantic segmentation task (Fig.~\ref{fig:teaser}, bottom left). With \emph{PT$_{K\%}$+FT$_{K\%}$}, Panda surpasses both the prior SSL baseline (PoLAr-MAE~\cite{polarmae}) and fully supervised UResNet~\cite{choy20194dspatiotemporalconvnetsminkowski,domine2020scalable} at low labels; pre-training and fine-tuning on the same 100 events results in performance comparable to training the UResNet on $>10,000$ images (85.2\%), and doing the same with just 10,000 images results in SOTA performance (97.7\%). Additionally, Panda greatly surpasses the supervised baselines at full label budget. With \emph{PT$_{100\%}$+FT$_{K\%}$}, a single backbone pretrained on 1M unlabeled events yields consistent gains for all $K$, comparable to the previous SOTA supervised baseline even at $0.1\%$ labels (-0.2\%)--a 1,000$\times$ increase in data-efficiency. Figure~\ref{fig:semseg_scaling} visualizes this trend: Panda (full fine-tuning) dominates the curve, with a from-scratch PTv3~\cite{wu2024pointtransformerv3simpler} closing the gap at high data counts, and the previous baseline UResNet being greatly outperformed by both at all dataset sizes. 

\paragraph{Semantic segmentation parameter efficiency.}
In Table~\ref{tab:semseg_high_stats}, we show the parameter efficiency of Panda when probed or fine-tuned with $K\!=\!100\%$ on the semantic segmentation task. We additionally include per-class $F_1$. Linear probing on frozen Panda features attains high ($>$94\%) $F_1$ across all classes except Delta rays with no backbone updates, and a small multi-scale decoder on top of frozen features matches full fine-tuning within a narrow (0.2\%) margin. Besides the delta ray result, where full-fine-tuning improved +15.4\% over linear probing, this indicates that most of the pixel-level signal is already present in the pretrained representation.

\begin{figure}[t]
    \centering
    \vspace{-5mm}
    \includegraphics[width=0.9\linewidth]{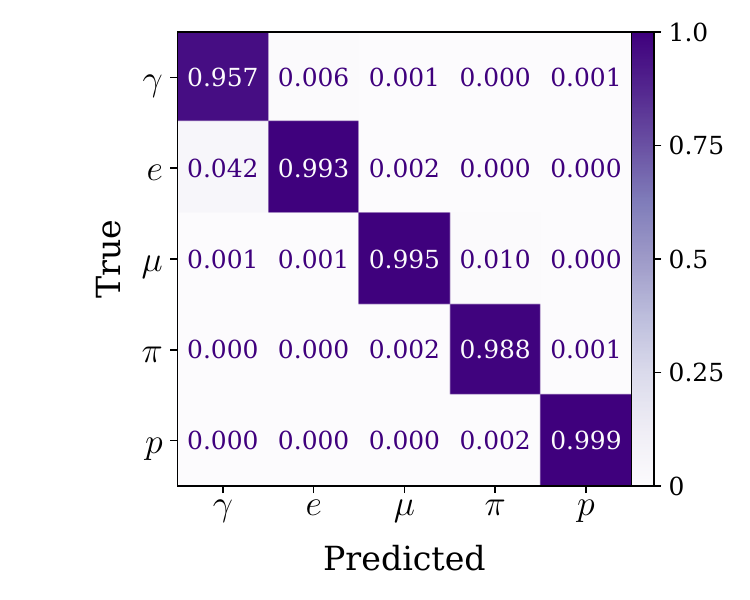}
    \vspace{-3mm}
    \caption{\textbf{Confusion matrix for particle classification, normalized by column.} The entries only include particles matched via IoU. The diagonals represent precision. Most confusion stems from the difficulty of $\gamma/e$ and $\mu/\pi$ discrimination.}
    \label{fig:confusion}
\end{figure}

\paragraph{Panoptic segmentation data efficiency.}
In Table~\ref{tab:insseg_data_eff} we assess data efficiency for panoptic segmentation at the particle and interaction level. We show that PQ and ARI increase monotonically with label count and remain relatively high under limited supervision (E.g., PQ/ARI of $(82.6\%,95.6\%)$ at $K=1\%$ for particle-level segmentation). Full fine-tuning achieves mean PQ/ARI of (92.5\%/98.0\%) for particle grouping and (97.6\%/94.9\%) for interaction grouping. For context, SPINE~\cite{drielsma2021scalableendtoenddeeplearningbaseddata} reports \(98.2\%\) ARI for particle clustering on PILArNet~\cite{adams2020pilarnetpublicdatasetparticle}; Panda reaches \(98.0\%\) with full fine-tuning and \(97.3\%\) with only the 4M decoder trained, indicating that a single pretrained sensor-level backbone plus a compact head is competitive with highly specialized pipelines.

\begin{figure}[t]
    \centering
\includegraphics[width=1\linewidth]{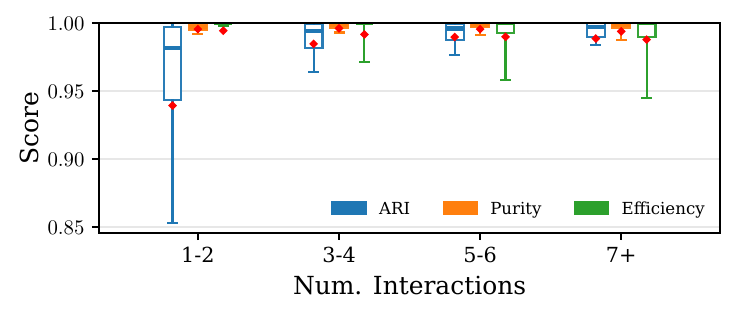}
    \vspace{-8mm}
    \caption{\textbf{Per-interaction panoptic segmentation across image densities.} We look at box plots of ARI, purity, and efficiency statistics over 10,000 images as a function of the number of interactions within each image. ARI appears to improve as a function of interaction number, while purity and efficiency decrease.}
    \label{fig:ari_grouped_by_num_interactions_boxplot_vtx}
\end{figure}

\paragraph{Panoptic segmentation parameter efficiency.}
In Table~\ref{tab:insseg_high_stats}, we show the parameter efficiency of Panda when fine-tuned either fully or just with the 4M decoder with $K\!=\!100\%$ (1M images) on the panoptic segmentation task. Here, full fine-tuning of the Panda backbone and set-prediction decoder outperforms the same architecture trained from scratch under identical training budgets (+1.4\% ARI for particle-level, +1.2\% ARI for interaction-level). We additionally include per-class ARI scores. We find that training just the decoder performs within (0.7\%/0.4\%) ARI of results of full fine-tuning, indicating that the representations learned by Panda indeed encodes particle- and interaction-instance-level information. We also evaluate Panda on 10,000 held-out images, and in Fig.~\ref{fig:ari_summary_boxplot} provide histograms of the ARI distributions across total images, and individual particle types. We include two additional metrics, purity and efficiency, which are described in \cite{koh2020scalableproposalfreeinstancesegmentation}. Within the histograms are box plots for ARI, purity, and efficiency. The red diamonds represent the means, the lines within the boxes the medians, the boxes the interquartile ranges (IQR), and the whiskers the 10th and 90th percentiles. These IQR are presented in parenthesis next to each box plot for easy reading.  We additionally provide a confusion matrix for matched particle predictions in Fig.~\ref{fig:confusion}, which shows a mean precision of 98.6\% for particle recognition. In Fig.~\ref{fig:ari_grouped_by_num_interactions_boxplot_vtx}, we provide box plots of ARI, purity, and efficiency statistics for interaction-level segmentation as a function of the number of interactions within each image. We fine that ARI appears to improve as a function of interaction number, while purity and efficiency decrease.

\begin{figure}
    \centering
    \includegraphics[width=1.0\linewidth]{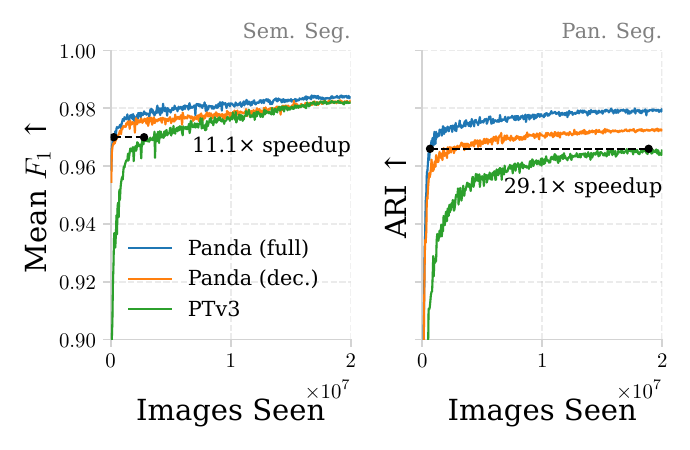}
    \vspace{-5mm}
    \caption{\textbf{Performance metrics over the course of fine-tuning.} We compare Panda fine-tuned by training all parameters (\textcolor{C0}{blue}), just the decoder (\textcolor{C1}{orange}), or completely from scratch without pre-training (\textcolor{C2}{green}). Fully fine-tuning Panda reaches 97\% mean $F_1$ and 96.6\% ARI in $11.1\times$ and $29.1\times$ faster than training the same model from scratch.}
    \label{fig:metric_vs_img}
\end{figure}

\paragraph{Transfer efficiency.}
In Fig.~\ref{fig:metric_vs_img}, we compare the validation $F_1$ and ARI  over the length of fine-tuning for semantic and particle-level panoptic segmentation. Panda reaches $\sim$97\% mean $F_1$ and ARI far earlier than from-scratch training, requiring {11.1$\times$} less images for semantic and {29.1$\times$} less images for particle-level panoptic segmentation. We also see a pronounced divergence in results for panoptic segmentation, which we attribute to the fact that set-prediction models notoriously take many iterations to converge~\cite{m2f}. We hypothesize that given enough training time, a model trained from scratch would catch up to full fine-tuning. This nonetheless shows the transfer power that a generally reusable backbone provides.

\begin{table}
    \centering
        \tablestyle{1.0pt}{1.08}
        \input{table/insseg_data_efficiency}
        \vspace{-3mm}
        \captionof{table}{\textbf{Data efficiency -- panoptic segmentation.} Here, the Panda backbone is frozen with just the set-prediction module trainable.}
        \label{tab:insseg_data_eff}
\end{table}

\begin{table}
    \centering
        \tablestyle{1.0pt}{1.08}
        \input{table/insseg_high_stats}
        \vspace{-3mm}
        \captionof{table}{\textbf{Parameter efficiency -- panoptic segmentation.} We fine-tune Panda pre-trained on 100\% of the dataset (1M images).}
        \label{tab:insseg_high_stats}
        \vspace{1mm}
    \vspace{1mm}
\end{table}

%% file: table/insseg_data_efficiency.tex
\begin{tabular}{lcccccccccc}
\toprule
Instance Seg. & \multicolumn{5}{c}{PQ} & \multicolumn{5}{c}{ARI} \\
 \cmidrule(lr){1-1} \cmidrule(lr){2-6} \cmidrule(lr){7-11} 
Task & 0.01\% & 0.1\% & 1\% & 10\% & 100\% & 0.01\% & 0.1\% & 1\% & 10\% & 100\% \\
\midrule
Particle Id. & 42.5 & 67.1 & \cellcolor[HTML]{F4F4FF}82.6 & \cellcolor[HTML]{E3E3FF}88.1 & \cellcolor[HTML]{D6D6FF}\textbf{89.5} & 73.4 & 91.4 & \cellcolor[HTML]{F4F4FF}95.6 & \cellcolor[HTML]{E3E3FF}97.1 & \cellcolor[HTML]{D6D6FF}\textbf{97.3} \\
Interaction Id. & 53.9 & 71.0 & \cellcolor[HTML]{F4F4FF}91.8 & \cellcolor[HTML]{E3E3FF}93.8 & \cellcolor[HTML]{D6D6FF}\textbf{96.6} & 68.3 & 81.6 & \cellcolor[HTML]{F4F4FF}92.1 & \cellcolor[HTML]{E3E3FF}93.3 & \cellcolor[HTML]{D6D6FF}\textbf{94.5} \\
\bottomrule
\end{tabular}

%% file: table/insseg_high_stats.tex
\begin{tabular}{lrcc@{\hspace{0.5\tabcolsep}}c@{\hspace{0.5\tabcolsep}}ccccccc}
\toprule
Instance Seg. & \multicolumn{1}{c}{Param.} & \multicolumn{2}{c}{Interaction Id.} & ~~ & \multicolumn{7}{c}{Particle Identification} \\
  \cmidrule(lr){1-1}   \cmidrule(lr){2-2} \cmidrule(lr){3-4} \cmidrule(lr){6-12}
Method & & PQ & ARI & & PQ & ARI & $\gamma$ & $e$ & $\mu$ & $\pi$ & $p$ \\
\midrule
\multicolumn{12}{l}{\textit{Supervised}} \\
\thiswork~PTv3~\cite{wu2024pointtransformerv3simpler} & 95.1~M & \cellcolor[HTML]{F4F4FF}~96.3 & \cellcolor[HTML]{F4F4FF}93.7 & & \cellcolor[HTML]{F4F4FF}86.9 & \cellcolor[HTML]{F4F4FF}96.6 & \cellcolor[HTML]{F4F4FF}93.5 & \cellcolor[HTML]{F4F4FF}94.9 & \cellcolor[HTML]{F4F4FF}98.6 & \cellcolor[HTML]{F4F4FF}94.2 & \cellcolor[HTML]{F4F4FF}98.1 \\
\midrule
\multicolumn{11}{l}{\textit{Self-supervised}} \\
\thiswork~Panda (dec.) & 4.4~M & \cellcolor[HTML]{E3E3FF}~96.6 & \cellcolor[HTML]{E3E3FF}94.5 & & \cellcolor[HTML]{E3E3FF}89.5 & \cellcolor[HTML]{E3E3FF}97.3 & \cellcolor[HTML]{E3E3FF}96.2 & \cellcolor[HTML]{E3E3FF}96.3 & \cellcolor[HTML]{E3E3FF}99.1 & \cellcolor[HTML]{E3E3FF}95.7 & \cellcolor[HTML]{E3E3FF}98.4 \\
\thiswork~Panda (full) & 95.1~M & \cellcolor[HTML]{D6D6FF}~\textbf{97.6} & \cellcolor[HTML]{D6D6FF}\textbf{94.9} & & \cellcolor[HTML]{D6D6FF}\textbf{92.5} & \cellcolor[HTML]{D6D6FF}\textbf{98.0} & \cellcolor[HTML]{D6D6FF}\textbf{98.4} & \cellcolor[HTML]{D6D6FF}\textbf{97.2} & \cellcolor[HTML]{D6D6FF}\textbf{99.3} & \cellcolor[HTML]{D6D6FF}\textbf{96.0} & \cellcolor[HTML]{D6D6FF}\textbf{98.6} \\
\bottomrule
\multicolumn{12}{r}{\thiswork~This work}  \\
\end{tabular}

%% file: secs/5_conclusion.tex
\section{Discussion and Conclusion}

LArTPC images lend themselves to self-supervised soft-clustering methods that have been pervasive in natural image and point cloud understanding. We show this by pre-training Panda, a hierarchical point cloud encoder, via self-distillation on 100–1,000,000 unlabeled LArTPC images and fine-tuning on several disparate reconstruction tasks that probe the model's internal understanding of particle physics.

We show that the representations learned are indeed semantically rich and encode physically meaningful information that allows for extreme data efficiency when transferring to downstream tasks like semantic segmentation and particle/interaction identification. We first show on the semantic segmentation task--a crucial initial step in the usual multi-level reconstruction algorithms for LArTPC images--that Panda greatly outperforms the previous state-of-the-art Sparse UResNet across all dimensions, and show that pre-training specifically results in massive increases in performance over the UResNet in the low-data regime. This shows that there is a path to training exceptionally successful models with just a few examples, labeled either from simulation or by hand, thereby reducing the endemic burden of simulation calibration within experiments. 

Beyond task-specific performance, we show that a small set-prediction head, just 5\% of the size of the backbone encoder, can be used to perform full particle identification comparably to SPINE~\cite{drielsma2021scalableendtoenddeeplearningbaseddata}, the current SOTA in the task, without any hand-crafted priors or information bottlenecks. We show that it's also possible to distinguish different particle interactions within an image from one another, showing that Panda encodes a sense of causality into its representations.

Relative to other attention-based encoders like PoLAr-MAE~\cite{polarmae}, our hierarchical design mitigates the quadratic cost of global attention but does not fully resolve it; state-space models~\cite{gu2022efficientlymodelinglongsequences, gu2024mambalineartimesequencemodeling} (e.g. Mamba~\cite{gu2024mambalineartimesequencemodeling}-style architectures) may offer different accuracy--efficiency trade-offs, especially for edge deployments or accelerators with constrained memory hierarchies.

Overall, Panda moves toward general sensor-level models that learn from raw measurements to support coherent, event-level understanding of physical processes. We hope this work encourages more research into AI for science utilizing raw data products. Future directions are described in the supplementary.

%% file: secs/7_acknow.tex
\section{Acknowledgements}

This work is supported by the U.S. Department of Energy, Office of Science, and Office of High Energy Physics under Contract No. DE-AC02-76SF00515.

%% file: secs/6_suppl.tex
\clearpage
\appendix

\section{Limitations and further work}

\paragraph{Simulated data.}
Our study evaluates Panda exclusively on simulated LArTPC data from a single detector configuration. In practice, deploying a sensor-level foundation model will require pre-training and fine-tuning on real data, together with detailed validation against well understood control samples such as through-going muons, Michel electrons, and stopping protons~\cite{Adams_2020_0,Acciarri_2017_0}. Establishing such a program is an experiment-scale effort and is therefore outside the scope of this work. 

\paragraph{Scope of reconstruction.}
Panda currently supports dense reconstruction of particles and interactions, but it (along with previous ML-based reconstruction algorithms) \textit{does not} perform the full image reconstruction with analysis-ready physics artifacts. Downstream physics analyses require per-instance kinematic quantities such as energies, momenta, directions, vertex positions, and associated uncertainties. Currently, this is accomplished by taking the outputs of e.g. Pandora~\cite{Acciarri2018} or SPINE~\cite{drielsma2021scalableendtoenddeeplearningbaseddata} and applying hand-crafted algorithms that exploit physics knowledge by, for example, comparing energy deposition distribution over the length of the particle track ($dE/dx$) to well-known theoretical distributions. In essence, this is pattern recognition, and so it is perhaps reasonable to suggest that machine learning algorithms may be able to perform this analysis as well. The set-prediction head we introduce is naturally suited for regression of per-instance properties; for example, one can attach small heads that predict momentum for particles or initial vertex positions for predicted interactions. Whether the representations learned by Panda can reliably predict these properties is an interesting question for further work.

\paragraph{Multi-modality.}
The current model operates on 3D charge images that are partially processed: in the pixelated LArTPC \cite{larpix} case, we are effectively using charge after signal deconvolution; in the context of wire-plane LArTPCs, we use 3D volumes reconstructed from three 2D views~\cite{Qian_2018}. Many LArTPCs are also equipped with photomultiplier tubes (PMTs) or other light sensors that record prompt scintillation light with fine time resolution and coarse spatial information. PMT waveforms are used in existing reconstruction chains for triggering~\cite{Kaleko_2013}, cosmic-ray rejection~\cite{Heggestuen_2024}, and calorimetry at low energies~\cite{Foreman_2020}. Extending Panda-style pre-training to jointly model charge and light, either through a shared backbone or coordinated encoders for PMT, TPC, and 2D wire-plane views could improve timing, background rejection, and low-energy response.

\paragraph{Architectural and computational constraints.}
Our backbone is a large attention-based point transformer with a hierarchical encoder that achieves strong performance but carries a substantial training and inference cost. While the hierarchical design mitigates the quadratic cost of global attention, it does not eliminate it; scaling to larger volumes and higher occupancies may be challenging for experiments with limited compute. Exploring more efficient sequence models such as state-space or hybrid convolution–state-space architectures~\cite{gu2024mambalineartimesequencemodeling,gu2022efficientlymodelinglongsequences,poli2023hyenahierarchylargerconvolutional,ku2025systemsalgorithmsconvolutionalmultihybrid}, as well as model compression and distillation for edge deployment, is a natural extension.

\paragraph{Next steps.}
Many potential uses of sensor-level representations remain unexplored. Promising directions include few-shot adaptation in settings with limited or poorly calibrated simulation for tasks such as track--shower separation, Michel electron tagging, and particle identification; content-based event retrieval and anomaly detection, for example querying ``find events like this'' in large archives of LArTPC data; and human-in-the-loop interfaces where language models reason over Panda features and natural-language annotations, enabling interactive labeling tools or captioning of events that lower the barrier to entry for students and non-experts. Additionally, building multi-experiment training corpora that span detectors and operating conditions may further drive detector-invariant embeddings that support rapid adaptation to new experiments.

\section{Examples}
\subsection{Semantic segmentation}
\begin{figure*}
    \centering
    \includegraphics[width=0.9\linewidth]{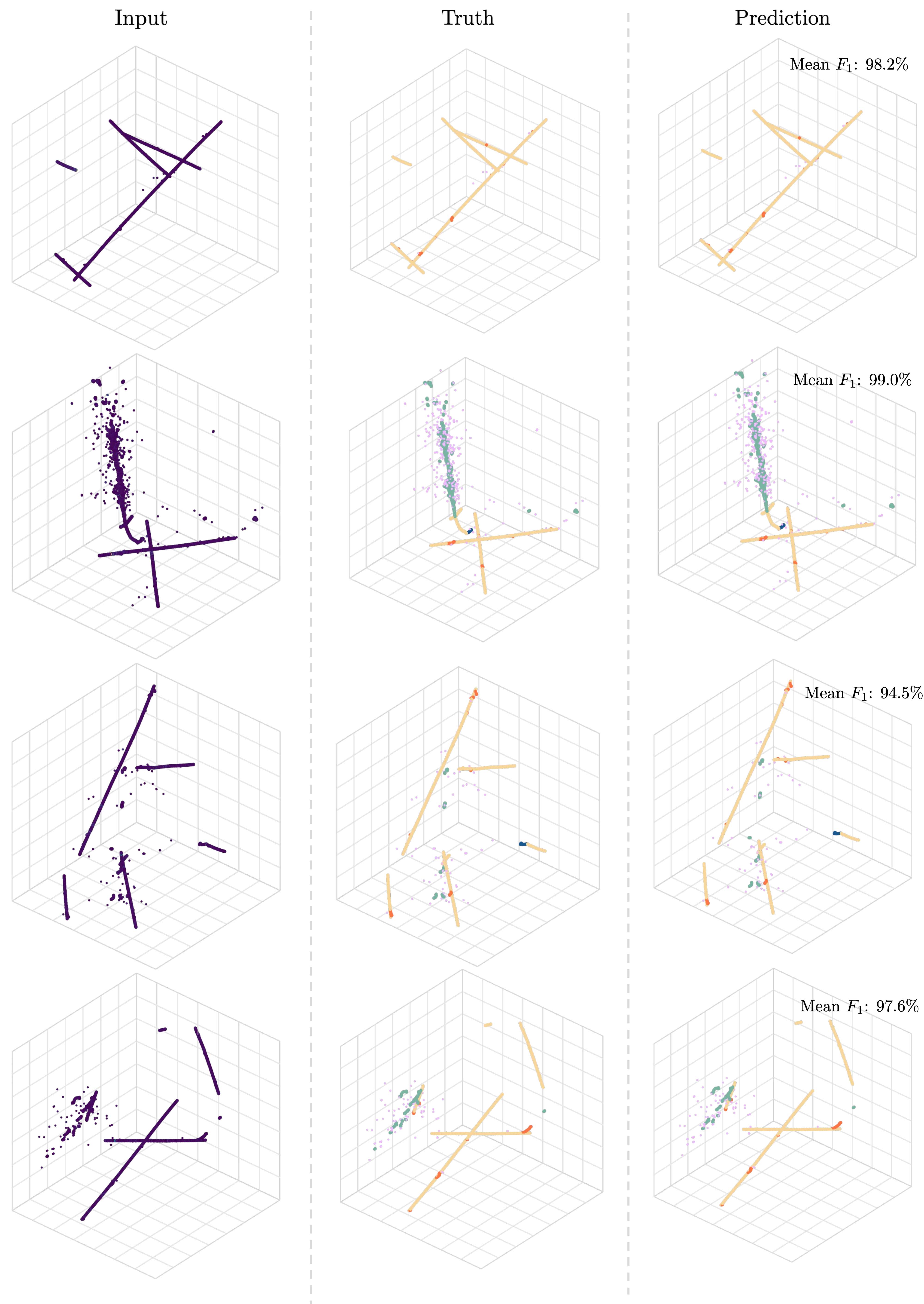}
    \caption{\textbf{Semantic segmentation examples}}
    \label{fig:semseg_ex}
\end{figure*}

We provide randomly sampled semantic segmentation examples along with $F_1$ scores in Fig.~\ref{fig:semseg_ex}.

\subsection{Panoptic segmentation}

\subsubsection{Particle-level}

\begin{figure*}
    \centering
    \includegraphics[width=0.9\linewidth]{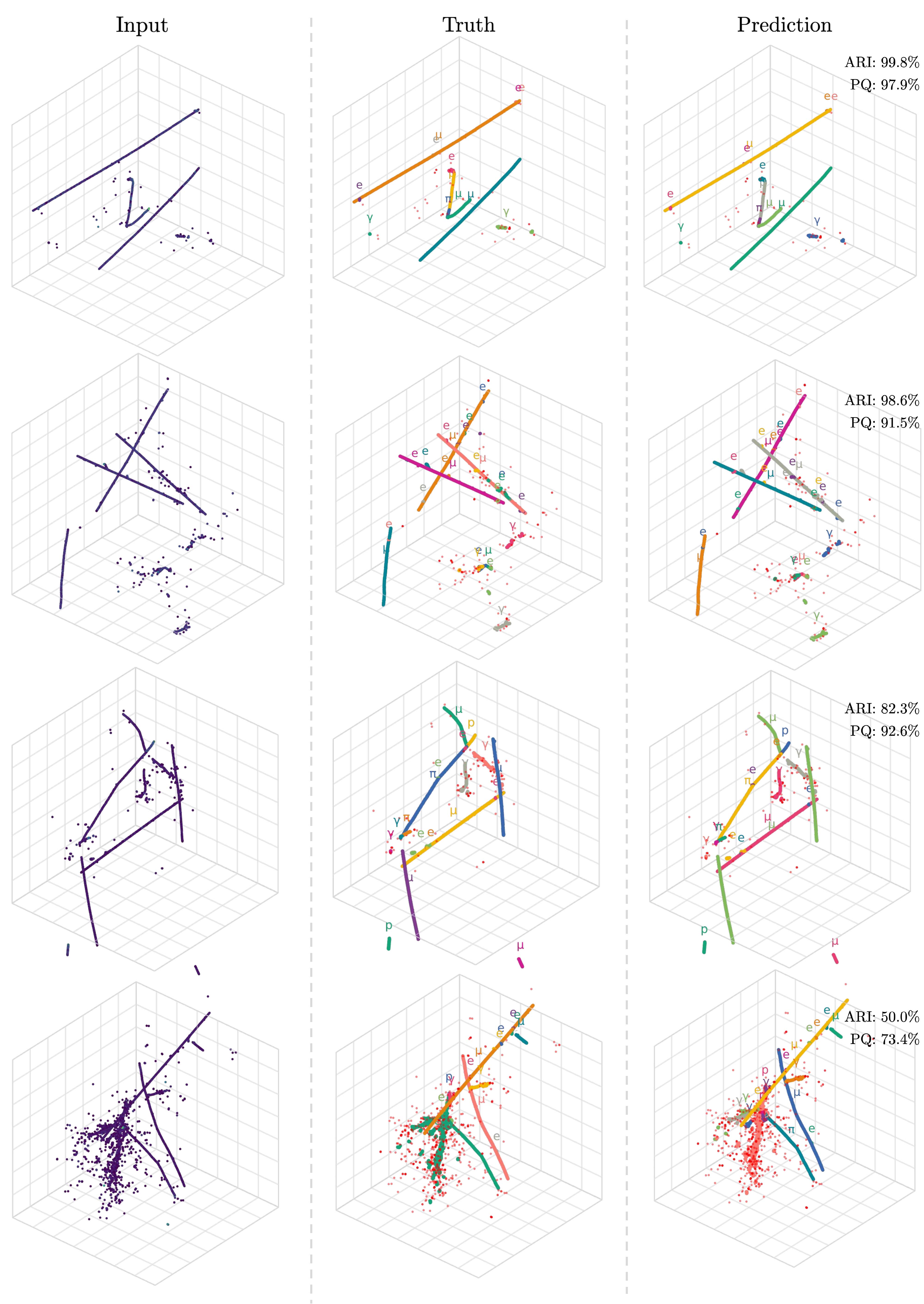}
    \caption{\textbf{Particle identification examples.} We provide four examples (two random, two poor) of Panda's particle-level event reconstruction.}
    \label{fig:pid_insseg_ex}
\end{figure*}

Refer to Fig.~\ref{fig:pid_insseg_ex} for two random and two poorly reconstructed images drawn from 1,000 images in the test set. We point out the last example, our worst example, which shows a possible labeling error in the PILArNet-M dataset. Here, what is labeled in the truth image is an electron actually contains a number of non-EM particles, as evident with the track-like trajectory. Panda sees this track-like particle and labels it as such.

\subsubsection{Interaction-level}

\begin{figure*}[th]
    \centering
\includegraphics[width=0.9\linewidth]{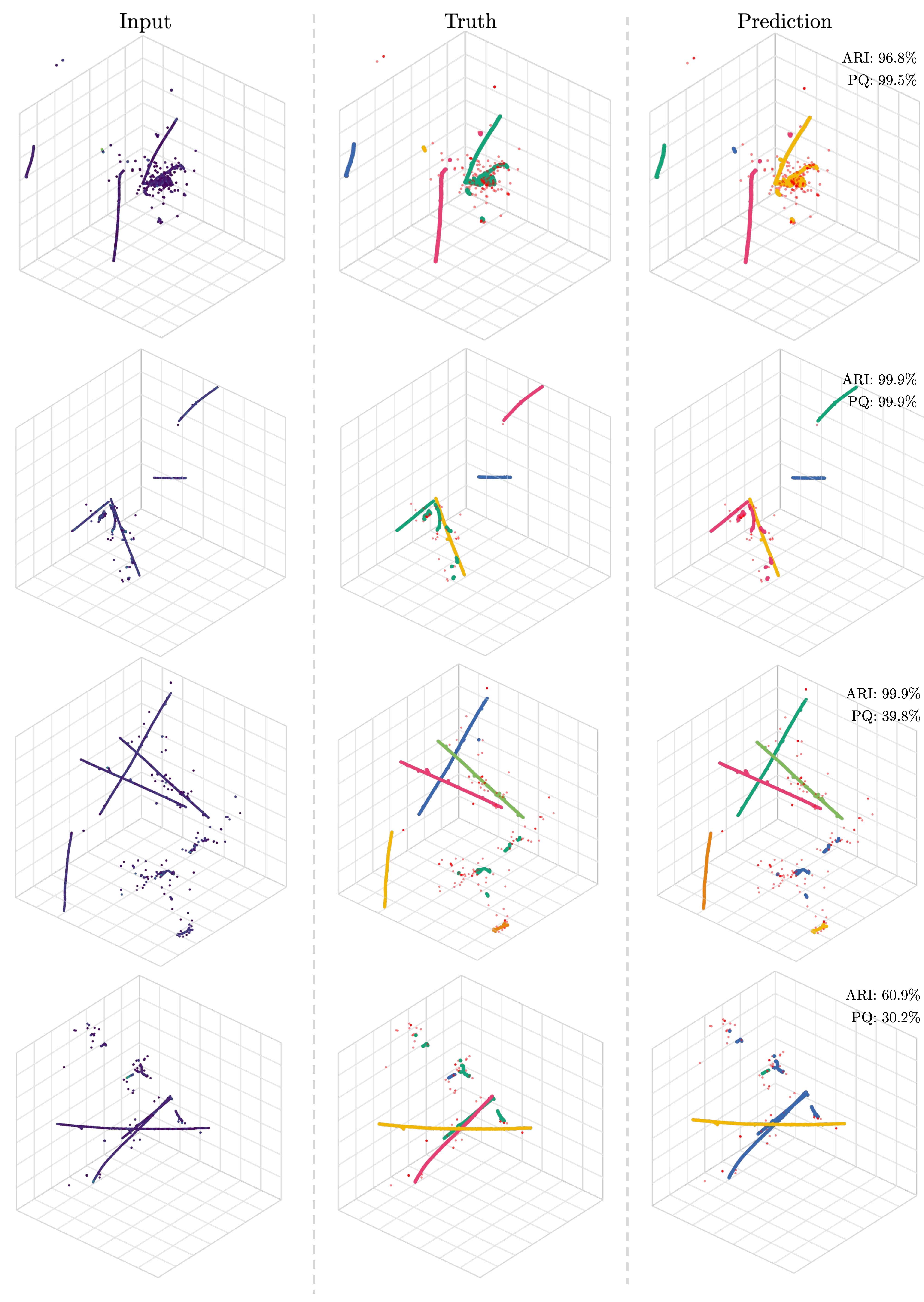}
    \caption{\textbf{Interaction identification examples} We provide four examples (two random, two poor) of Panda's interaction-level event reconstruction.}
    \label{fig:intn_insseg_ex}
\end{figure*}

Refer to Fig.~\ref{fig:intn_insseg_ex} for two random and two poorly reconstructed images drawn from 1,000 images in the test set.

\section{Computational Considerations}

\begin{figure}
        \centering
        \includegraphics[width=\linewidth]{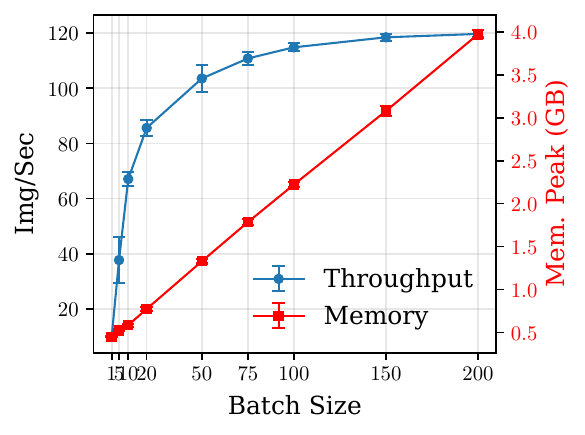}
        \caption{\textbf{Semantic segmentation throughput (\textcolor{C0}{blue}) and peak memory usage (\textcolor{red}{red}) as a function of batch size.} Markers indicate means, and error bars correspond to the standard deviation across 5 forward passes.}
        \label{fig:segsem_comp}
\end{figure}

\begin{figure}
        \centering
        \includegraphics[width=\linewidth]{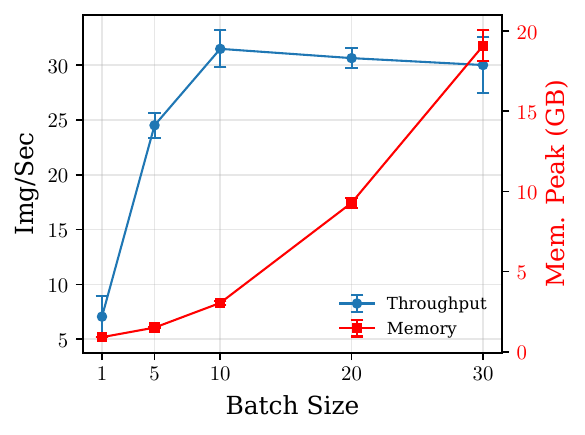}
        \caption{\textbf{Panoptic segmentation throughput (\textcolor{C0}{blue}) and peak memory usage (\textcolor{red}{red}) as a function of batch size.} Markers indicate means, and error bars correspond to the standard deviation across 5 forward passes.}
        \label{fig:insseg_comp}
\end{figure}

Figures~\ref{fig:segsem_comp} and \ref{fig:insseg_comp} report throughput (images per second) and peak GPU memory usage as a function of batch size for semantic and panoptic segmentation. At peak throughput, the semantic segmentation model processes 112.4\,img/s with a peak VRAM usage of 2.2\,GB. The panoptic segmentation model processes 31.5\,img/s with a peak VRAM usage of 3.0\,GB. This corresponds to 9.7M images per A100-day to semantic segmentation and 2.7M images per A100-day.

%% file: main.bib
@article{Abratenko2023,
  title = {ICARUS at the Fermilab Short-Baseline Neutrino program: initial operation},
  volume = {83},
  ISSN = {1434-6052},
  url = {http://dx.doi.org/10.1140/epjc/s10052-023-11610-y},
  DOI = {10.1140/epjc/s10052-023-11610-y},
  number = {6},
  journal = {The European Physical Journal C},
  publisher = {Springer Science and Business Media LLC},
  author = {Abratenko,  P. and Aduszkiewicz,  A. and Akbar,  F. and Pons,  M. Artero and Asaadi,  J. and Aslin,  M. and Babicz,  M. and Badgett,  W. F. and Bagby,  L. F. and Baibussinov,  B. and Behera,  B. and Bellini,  V. and Beltramello,  O. and Benocci,  R. and Berger,  J. and Berkman,  S. and Bertolucci,  S. and Bertoni,  R. and Betancourt,  M. and Bettini,  M. and Biagi,  S. and Biery,  K. and Bitter,  O. and Bonesini,  M. and Boone,  T. and Bottino,  B. and Braggiotti,  A. and Brailsford,  D. and Bremer,  J. and Brice,  S. J. and Brio,  V. and Brizzolari,  C. and Brown,  J. and Budd,  H. S. and Calaon,  F. and Campani,  A. and Carber,  D. and Carneiro,  M. and Terrazas,  I. Caro and Carranza,  H. and Casazza,  D. and Castellani,  L. and Castro,  A. and Centro,  S. and Cerati,  G. and Chalifour,  M. and Chambouvet,  P. and Chatterjee,  A. and Cherdack,  D. and Cherubini,  S. and Chithirasreemadam,  N. and Cicerchia,  M. and Cicero,  V. and Coan,  T. and Cocco,  A. G. and Convery,  M. R. and Copello,  S. and Cristaldo,  E. and Dange,  A. A. and de Icaza Astiz,  I. and De Roeck,  A. and Di Domizio,  S. and Di Noto,  L. and Di Stefano,  C. and Di Ferdinando,  D. and Diwan,  M. and Dolan,  S. and Domine,  L. and Donati,  S. and Doubnik,  R. and Drielsma,  F. and Dyer,  J. and Dytman,  S. and Fabre,  C. and Fabris,  F. and Falcone,  A. and Farnese,  C. and Fava,  A. and Ferguson,  H. and Ferrari,  A. and Ferraro,  F. and Gallice,  N. and Garcia,  F. G. and Geynisman,  M. and Giarin,  M. and Gibin,  D. and Gigli,  S. G. and Gioiosa,  A. and Gu,  W. and Guerzoni,  M. and Guglielmi,  A. and Gurung,  G. and Hahn,  S. and Hardin,  K. and Hausner,  H. and Heggestuen,  A. and Hilgenberg,  C. and Hogan,  M. and Howard,  B. and Howell,  R. and Hrivnak,  J. and Iliescu,  M. and Ingratta,  G. and James,  C. and Jang,  W. and Jung,  M. and Jwa,  Y.-J. and Kashur,  L. and Ketchum,  W. and Kim,  J. S. and Koh,  D.-H. and Kose,  U. and Larkin,  J. and Laurenti,  G. and Lukhanin,  G. and Marchini,  S. and Marshall,  C. M. and Martynenko,  S. and Mauri,  N. and Mazzacane,  A. and McFarland,  K. S. and Méndez,  D. P. and Menegolli,  A. and Meng,  G. and Miranda,  O. G. and Mladenov,  D. and Mogan,  A. and Moggi,  N. and Montagna,  E. and Montanari,  C. and Montanari,  A. and Mooney,  M. and Moreno-Granados,  G. and Mueller,  J. and Naples,  D. and Nebot-Guinot,  M. and Nessi,  M. and Nichols,  T. and Nicoletto,  M. and Norris,  B. and Palestini,  S. and Pallavicini,  M. and Paolone,  V. and Papaleo,  R. and Pasqualini,  L. and Patrizii,  L. and Peghin,  R. and Petrillo,  G. and Petta,  C. and Pia,  V. and Pietropaolo,  F. and Poirot,  J. and Poppi,  F. and Pozzato,  M. and Prata,  M. C. and Prosser,  A. and Putnam,  G. and Qian,  X. and Rampazzo,  G. and Rappoldi,  A. and Raselli,  G. L. and Rechenmacher,  R. and Resnati,  F. and Ricci,  A. M. and Riccobene,  G. and Rice,  L. and Richards,  E. and Rigamonti,  A. and Rosenberg,  M. and Rossella,  M. and Rubbia,  C. and Sala,  P. and Sapienza,  P. and Savage,  G. and Scaramelli,  A. and Scarpelli,  A. and Schmitz,  D. and Schukraft,  A. and Sergiampietri,  F. and Sirri,  G. and Smedley,  J. S. and Soha,  A. K. and Spanu,  M. and Stanco,  L. and Stewart,  J. and Suarez,  N. B. and Sutera,  C. and Tanaka,  H. A. and Tenti,  M. and Terao,  K. and Terranova,  F. and Togo,  V. and Torretta,  D. and Torti,  M. and Tortorici,  F. and Tosi,  N. and Tsai,  Y.-T. and Tufanli,  S. and Turcato,  M. and Usher,  T. and Varanini,  F. and Ventura,  S. and Vercellati,  F. and Vicenzi,  M. and Vignoli,  C. and Viren,  B. and Warner,  D. and Williams,  Z. and Wilson,  R. J. and Wilson,  P. and Wolfs,  J. and Wongjirad,  T. and Wood,  A. and Worcester,  E. and Worcester,  M. and Wospakrik,  M. and Yu,  H. and Yu,  J. and Zani,  A. and Zatti,  P. G. and Zennamo,  J. and Zettlemoyer,  J. C. and Zhang,  C. and Zucchelli,  S. and Zuckerbrot,  M.},
  year = {2023},
  month = jun 
}

@article{htfm-tbdq,
  title = {Enhancing DUNE's Solar Neutrino Capabilities with Neutral-Current Detection},
  author = {Meighen-Berger, Stephan A. and Newstead, Jayden L. and Beacom, John F. and Bell, Nicole F. and Dolan, Matthew J.},
  journal = {Phys. Rev. Lett.},
  volume = {135},
  issue = {1},
  pages = {011803},
  numpages = {10},
  year = {2025},
  month = {Jul},
  publisher = {American Physical Society},
  doi = {10.1103/htfm-tbdq},
  url = {https://link.aps.org/doi/10.1103/htfm-tbdq}
}

@article{PhysRevD.101.052001,
  title = {Search for heavy neutral leptons decaying into muon-pion pairs in the MicroBooNE detector},
  author = {Abratenko, P. and Alrashed, M. and An, R. and Anthony, J. and Asaadi, J. and Ashkenazi, A. and Balasubramanian, S. and Baller, B. and Barnes, C. and Barr, G. and Basque, V. and Berkman, S. and Bhanderi, A. and Bhat, A. and Bishai, M. and Blake, A. and Bolton, T. and Camilleri, L. and Caratelli, D. and Caro Terrazas, I. and Castillo Fernandez, R. and Cavanna, F. and Cerati, G. and Chen, Y. and Church, E. and Cianci, D. and Cohen, E. O. and Conrad, J. M. and Convery, M. and Cooper-Troendle, L. and Crespo-Anad\'on, J. I. and Del Tutto, M. and Devitt, A. and Domine, L. and Duffy, K. and Dytman, S. and Eberly, B. and Ereditato, A. and Escudero Sanchez, L. and Evans, J. J. and Fitzpatrick, R. S. and Fleming, B. T. and Foppiani, N. and Franco, D. and Furmanski, A. P. and Garcia-Gamez, D. and Gardiner, S. and Genty, V. and Goeldi, D. and Gollapinni, S. and Goodwin, O. and Gramellini, E. and Green, P. and Greenlee, H. and Gu, L. and Gu, W. and Guenette, R. and Guzowski, P. and Hamilton, P. and Hen, O. and Hill, C. and Horton-Smith, G. A. and Hourlier, A. and Huang, E.-C. and Itay, R. and James, C. and Jan de Vries, J. and Ji, X. and Jiang, L. and Jo, J. H. and Johnson, R. A. and Joshi, J. and Jwa, Y.-J. and Karagiorgi, G. and Ketchum, W. and Kirby, B. and Kirby, M. and Kobilarcik, T. and Kreslo, I. and LaZur, R. and Lepetic, I. and Li, Y. and Lister, A. and Littlejohn, B. R. and Lockwitz, S. and Lorca, D. and Louis, W. C. and Luethi, M. and Lundberg, B. and Luo, X. and Marchionni, A. and Marcocci, S. and Mariani, C. and Marshall, J. and Martin-Albo, J. and Martinez Caicedo, D. A. and Mason, K. and Mastbaum, A. and McConkey, N. and Meddage, V. and Mettler, T. and Miller, K. and Mills, J. and Mistry, K. and Mogan, A. and Mohayai, T. and Moon, J. and Mooney, M. and Moore, C. D. and Mousseau, J. and Murrells, R. and Naples, D. and Neely, R. K. and Nienaber, P. and Nowak, J. and Palamara, O. and Pandey, V. and Paolone, V. and Papadopoulou, A. and Papavassiliou, V. and Pate, S. F. and Paudel, A. and Pavlovic, Z. and Piasetzky, E. and Porzio, D. and Prince, S. and Pulliam, G. and Qian, X. and Raaf, J. L. and Radeka, V. and Rafique, A. and Ren, L. and Rochester, L. and Rogers, H. E. and Ross-Lonergan, M. and Rudolf von Rohr, C. and Russell, B. and Scanavini, G. and Schmitz, D. W. and Schukraft, A. and Seligman, W. and Shaevitz, M. H. and Sharankova, R. and Sinclair, J. and Smith, A. and Snider, E. L. and Soderberg, M. and S\"oldner-Rembold, S. and Soleti, S. R. and Spentzouris, P. and Spitz, J. and Stancari, M. and John, J. St. and Strauss, T. and Sutton, K. and Sword-Fehlberg, S. and Szelc, A. M. and Tagg, N. and Tang, W. and Terao, K. and Thornton, R. T. and Toups, M. and Tsai, Y.-T. and Tufanli, S. and Uchida, M. A. and Usher, T. and Van De Pontseele, W. and Van de Water, R. G. and Viren, B. and Weber, M. and Wei, H. and Wickremasinghe, D. A. and Williams, Z. and Wolbers, S. and Wongjirad, T. and Woodruff, K. and Wospakrik, M. and Wu, W. and Yang, T. and Yarbrough, G. and Yates, L. E. and Zeller, G. P. and Zennamo, J. and Zhang, C.},
  collaboration = {The MicroBooNE Collaboration},
  journal = {Phys. Rev. D},
  volume = {101},
  issue = {5},
  pages = {052001},
  numpages = {11},
  year = {2020},
  month = {Mar},
  publisher = {American Physical Society},
  doi = {10.1103/PhysRevD.101.052001},
  url = {https://link.aps.org/doi/10.1103/PhysRevD.101.052001}
}

@article{Kelly_2022,
   title={DUNE atmospheric neutrinos: Earth tomography},
   volume={2022},
   ISSN={1029-8479},
   url={http://dx.doi.org/10.1007/JHEP05(2022)187},
   DOI={10.1007/jhep05(2022)187},
   number={5},
   journal={Journal of High Energy Physics},
   publisher={Springer Science and Business Media LLC},
   author={Kelly, Kevin J. and Machado, Pedro A. N. and Martinez-Soler, Ivan and Perez-Gonzalez, Yuber F.},
   year={2022},
   month=may }

@article{PhysRevLett.123.131803,
  title = {DUNE as the Next-Generation Solar Neutrino Experiment},
  author = {Capozzi, Francesco and Li, Shirley Weishi and Zhu, Guanying and Beacom, John F.},
  journal = {Phys. Rev. Lett.},
  volume = {123},
  issue = {13},
  pages = {131803},
  numpages = {7},
  year = {2019},
  month = {Sep},
  publisher = {American Physical Society},
  doi = {10.1103/PhysRevLett.123.131803},
  url = {https://link.aps.org/doi/10.1103/PhysRevLett.123.131803}
}

@article{PhysRevLett.123.081801,
  title = {Sub-GeV Atmospheric Neutrinos and $CP$ Violation in DUNE},
  author = {Kelly, Kevin J. and Machado, Pedro A. N. and Martinez-Soler, Ivan and Parke, Stephen J. and Perez-Gonzalez, Yuber F.},
  journal = {Phys. Rev. Lett.},
  volume = {123},
  issue = {8},
  pages = {081801},
  numpages = {6},
  year = {2019},
  month = {Aug},
  publisher = {American Physical Society},
  doi = {10.1103/PhysRevLett.123.081801},
  url = {https://link.aps.org/doi/10.1103/PhysRevLett.123.081801}
}

@article{PhysRevD.103.095012,
  title = {Prospects for detecting boosted dark matter in DUNE through hadronic interactions},
  author = {Berger, Joshua and Cui, Yanou and Graham, Mathew and Necib, Lina and Petrillo, Gianluca and Stocks, Dane and Tsai, Yun-Tse and Zhao, Yue},
  journal = {Phys. Rev. D},
  volume = {103},
  issue = {9},
  pages = {095012},
  numpages = {13},
  year = {2021},
  month = {May},
  publisher = {American Physical Society},
  doi = {10.1103/PhysRevD.103.095012},
  url = {https://link.aps.org/doi/10.1103/PhysRevD.103.095012}
}

@article{PhysRevD.111.092006,
  title = {Supernova pointing capabilities of DUNE},
  author = {Abed Abud, A. et al.},
  collaboration = {DUNE Collaboration},
  journal = {Phys. Rev. D},
  volume = {111},
  issue = {9},
  pages = {092006},
  numpages = {24},
  year = {2025},
  month = {May},
  publisher = {American Physical Society},
  doi = {10.1103/PhysRevD.111.092006},
  url = {https://link.aps.org/doi/10.1103/PhysRevD.111.092006},
}

@article{PhysRevD.107.112012,
  title = {Impact of cross-section uncertainties on supernova neutrino spectral parameter fitting in the Deep Underground Neutrino Experiment},
  author = {Abed Abud, A. et al.},
  collaboration = {The DUNE Collaboration},
  journal = {Phys. Rev. D},
  volume = {107},
  issue = {11},
  pages = {112012},
  numpages = {25},
  year = {2023},
  month = {Jun},
  publisher = {American Physical Society},
  doi = {10.1103/PhysRevD.107.112012},
  url = {https://link.aps.org/doi/10.1103/PhysRevD.107.112012}
}

@article{Abi_2021,
   title={Prospects for beyond the Standard Model physics searches at the Deep Underground Neutrino Experiment: DUNE Collaboration},
   volume={81},
   ISSN={1434-6052},
   url={http://dx.doi.org/10.1140/epjc/s10052-021-09007-w},
   DOI={10.1140/epjc/s10052-021-09007-w},
   number={4},
   journal={The European Physical Journal C},
   publisher={Springer Science and Business Media LLC},
   author={Abi, B. et al.},
   year={2021},
   month={apr} 
}

@misc{shutt2024gampix,
      title={GAMPix: a novel fine-grained, low-noise and ultra-low power pixelated charge readout for TPCs}, 
      author={Tom Shutt and Bahrudin Trbalic and Aldo Pena-Perez and Steffen Luitz and Mark Convery and Angelo Dragone and Lorenzo Rota and Dietrich R. Freytag and Dionisio Doering and Filippo Mele and Miriam Moore and Hiro Tanaka and Yun-Tse Tsai},
      year={2024},
      eprint={2402.00902},
      archivePrefix={arXiv},
      primaryClass={physics.ins-det},
      url={https://arxiv.org/abs/2402.00902}, 
}

@article{Abi_2021_sn,
   title={Supernova neutrino burst detection with the Deep Underground Neutrino Experiment: DUNE Collaboration},
   volume={81},
   ISSN={1434-6052},
   url={http://dx.doi.org/10.1140/epjc/s10052-021-09166-w},
   DOI={10.1140/epjc/s10052-021-09166-w},
   year = {2021},
   month = {may},
   number={5},
   journal={The European Physical Journal C},
   publisher={Springer Science and Business Media LLC},
   author={Abi, B. et al.}
}

@article{Acciarri_2017,
doi = {10.1088/1748-0221/12/02/P02017},
url = {https://dx.doi.org/10.1088/1748-0221/12/02/P02017},
year = {2017},
month = {feb},
publisher = {},
volume = {12},
number = {02},
pages = {P02017},
author = {Acciarri, R. et al.},
title = {Design and construction of the MicroBooNE detector},
journal = {Journal of Instrumentation}
}

@article{ARAMAKI2020107,
title = {Dual MeV gamma-ray and dark matter observatory - GRAMS Project},
journal = {Astroparticle Physics},
volume = {114},
pages = {107-114},
year = {2020},
issn = {0927-6505},
doi = {https://doi.org/10.1016/j.astropartphys.2019.07.002},
url = {https://www.sciencedirect.com/science/article/pii/S0927650519300805},
author = {Tsuguo Aramaki and Per Ola Hansson Adrian and Georgia Karagiorgi and Hirokazu Odaka}
}

@misc{sbn2015,
      title={A Proposal for a Three Detector Short-Baseline Neutrino Oscillation Program in the Fermilab Booster Neutrino Beam}, 
      author={R. Acciarri et al.},
      year={2015},
      eprint={1503.01520},
      archivePrefix={arXiv},
      primaryClass={physics.ins-det},
      url={https://arxiv.org/abs/1503.01520}, 
}

@misc{golling2024maskedparticlemodelingsets,
      title={Masked Particle Modeling on Sets: Towards Self-Supervised High Energy Physics Foundation Models}, 
      author={Tobias Golling and Lukas Heinrich and Michael Kagan and Samuel Klein and Matthew Leigh and Margarita Osadchy and John Andrew Raine},
      year={2024},
      eprint={2401.13537},
      archivePrefix={arXiv},
      primaryClass={hep-ph},
      url={https://arxiv.org/abs/2401.13537}, 
}

@misc{r3sl,
      title={Re-Simulation-based Self-Supervised Learning for Pre-Training Foundation Models}, 
      author={Philip Harris and Michael Kagan and Jeffrey Krupa and Benedikt Maier and Nathaniel Woodward},
      year={2024},
      eprint={2403.07066},
      archivePrefix={arXiv},
      primaryClass={hep-ph},
      url={https://arxiv.org/abs/2403.07066}, 
}

@misc{drielsma2021scalableendtoenddeeplearningbaseddata,
      title={Scalable, End-to-End, Deep-Learning-Based Data Reconstruction Chain for Particle Imaging Detectors}, 
      author={Francois Drielsma and Kazuhiro Terao and Laura Dominé and Dae Heun Koh},
      year={2021},
      eprint={2102.01033},
      archivePrefix={arXiv},
      primaryClass={hep-ex},
      url={https://arxiv.org/abs/2102.01033}, 
}

@misc{he2021maskedautoencodersscalablevision,
      title={Masked Autoencoders Are Scalable Vision Learners}, 
      author={Kaiming He and Xinlei Chen and Saining Xie and Yanghao Li and Piotr Dollár and Ross Girshick},
      year={2021},
      eprint={2111.06377},
      archivePrefix={arXiv},
      primaryClass={cs.CV},
      url={https://arxiv.org/abs/2111.06377}, 
}

@misc{pang2022maskedautoencoderspointcloud,
      title={Masked Autoencoders for Point Cloud Self-supervised Learning}, 
      author={Yatian Pang and Wenxiao Wang and Francis E. H. Tay and Wei Liu and Yonghong Tian and Li Yuan},
      year={2022},
      eprint={2203.06604},
      archivePrefix={arXiv},
      primaryClass={cs.CV},
      url={https://arxiv.org/abs/2203.06604}, 
}

@misc{yu2022pointbertpretraining3dpoint,
      title={Point-BERT: Pre-training 3D Point Cloud Transformers with Masked Point Modeling}, 
      author={Xumin Yu and Lulu Tang and Yongming Rao and Tiejun Huang and Jie Zhou and Jiwen Lu},
      year={2022},
      eprint={2111.14819},
      archivePrefix={arXiv},
      primaryClass={cs.CV},
      url={https://arxiv.org/abs/2111.14819}, 
}

@article{vaswani2017attention,
  title={Attention is all you need},
  author={Vaswani, A},
  journal={Advances in Neural Information Processing Systems},
  year={2017}
}

@misc{choy20194dspatiotemporalconvnetsminkowski,
      title={4D Spatio-Temporal ConvNets: Minkowski Convolutional Neural Networks}, 
      author={Christopher Choy and JunYoung Gwak and Silvio Savarese},
      year={2019},
      eprint={1904.08755},
      archivePrefix={arXiv},
      primaryClass={cs.CV},
      url={https://arxiv.org/abs/1904.08755}, 
}

@inproceedings{chen2020simple,
  title={A simple framework for contrastive learning of visual representations},
  author={Chen, Ting and Kornblith, Simon and Norouzi, Mohammad and Hinton, Geoffrey},
  booktitle={International conference on machine learning},
  pages={1597--1607},
  year={2020},
  organization={PMLR}
}

@inproceedings{he2020momentum,
  title={Momentum contrast for unsupervised visual representation learning},
  author={He, Kaiming and Fan, Haoqi and Wu, Yuxin and Xie, Saining and Girshick, Ross},
  booktitle={Proceedings of the IEEE/CVF conference on computer vision and pattern recognition},
  pages={9729--9738},
  year={2020}
}

@inproceedings{caron2021emerging,
  title={Emerging properties in self-supervised vision transformers},
  author={Caron, Mathilde and Touvron, Hugo and Misra, Ishan and J{\'e}gou, Herv{\'e} and Mairal, Julien and Bojanowski, Piotr and Joulin, Armand},
  booktitle={Proceedings of the IEEE/CVF international conference on computer vision},
  pages={9650--9660},
  year={2021}
}

@misc{zhou2022ibotimagebertpretraining,
      title={iBOT: Image BERT Pre-Training with Online Tokenizer}, 
      author={Jinghao Zhou and Chen Wei and Huiyu Wang and Wei Shen and Cihang Xie and Alan Yuille and Tao Kong},
      year={2022},
      eprint={2111.07832},
      archivePrefix={arXiv},
      primaryClass={cs.CV},
      url={https://arxiv.org/abs/2111.07832}, 
}

@techreport{rubbia1977liquid,
      author        = "Rubbia, Carlo",
      title         = "{The liquid-argon time projection chamber: a new concept
                       for neutrino detectors}",
      institution   = "CERN",
      reportNumber  = "CERN-EP-INT-77-8",
      address       = "Geneva",
      year          = "1977",
      url           = "https://cds.cern.ch/record/117852",
}

@misc{adams2020pilarnetpublicdatasetparticle,
      title={PILArNet: Public Dataset for Particle Imaging Liquid Argon Detectors in High Energy Physics}, 
      author={Corey Adams and Kazuhiro Terao and Taritree Wongjirad},
      year={2020},
      eprint={2006.01993},
      archivePrefix={arXiv},
      primaryClass={physics.ins-det},
      url={https://arxiv.org/abs/2006.01993}, 
}

@misc{lin2018focallossdenseobject,
      title={Focal Loss for Dense Object Detection}, 
      author={Tsung-Yi Lin and Priya Goyal and Ross Girshick and Kaiming He and Piotr Dollár},
      year={2018},
      eprint={1708.02002},
      archivePrefix={arXiv},
      primaryClass={cs.CV},
      url={https://arxiv.org/abs/1708.02002}, 
}

@article{Abi2020,
  title = {Volume IV. The DUNE far detector single-phase technology},
  volume = {15},
  ISSN = {1748-0221},
  url = {http://dx.doi.org/10.1088/1748-0221/15/08/T08010},
  DOI = {10.1088/1748-0221/15/08/t08010},
  number = {08},
  journal = {Journal of Instrumentation},
  publisher = {IOP Publishing},
  author = {Abi,  B. and Acciarri,  R. and Acero,  M.A. and Adamov,  G. and Adams,  D. and Adinolfi,  M. and Ahmad,  Z. and Ahmed,  J. and Alion,  T.},
  year = {2020},
  month = aug,
  pages = {T08010–T08010}
}

@misc{oquab2024dinov2learningrobustvisual,
      title={DINOv2: Learning Robust Visual Features without Supervision}, 
      author={Maxime Oquab and Timothée Darcet and Théo Moutakanni and Huy Vo and Marc Szafraniec and Vasil Khalidov and Pierre Fernandez and Daniel Haziza and Francisco Massa and Alaaeldin El-Nouby and Mahmoud Assran and Nicolas Ballas and Wojciech Galuba and Russell Howes and Po-Yao Huang and Shang-Wen Li and Ishan Misra and Michael Rabbat and Vasu Sharma and Gabriel Synnaeve and Hu Xu and Hervé Jegou and Julien Mairal and Patrick Labatut and Armand Joulin and Piotr Bojanowski},
      year={2024},
      eprint={2304.07193},
      archivePrefix={arXiv},
      primaryClass={cs.CV},
      url={https://arxiv.org/abs/2304.07193}, 
}

@misc{xie2020pointcontrastunsupervisedpretraining3d,
      title={PointContrast: Unsupervised Pre-training for 3D Point Cloud Understanding}, 
      author={Saining Xie and Jiatao Gu and Demi Guo and Charles R. Qi and Leonidas J. Guibas and Or Litany},
      year={2020},
      eprint={2007.10985},
      archivePrefix={arXiv},
      primaryClass={cs.CV},
      url={https://arxiv.org/abs/2007.10985}, 
}

@misc{wu2025sonataselfsupervisedlearningreliable,
      title={Sonata: Self-Supervised Learning of Reliable Point Representations}, 
      author={Xiaoyang Wu and Daniel DeTone and Duncan Frost and Tianwei Shen and Chris Xie and Nan Yang and Jakob Engel and Richard Newcombe and Hengshuang Zhao and Julian Straub},
      year={2025},
      eprint={2503.16429},
      archivePrefix={arXiv},
      primaryClass={cs.CV},
      url={https://arxiv.org/abs/2503.16429}, 
}

@misc{wu2024pointtransformerv3simpler,
      title={Point Transformer V3: Simpler, Faster, Stronger}, 
      author={Xiaoyang Wu and Li Jiang and Peng-Shuai Wang and Zhijian Liu and Xihui Liu and Yu Qiao and Wanli Ouyang and Tong He and Hengshuang Zhao},
      year={2024},
      eprint={2312.10035},
      archivePrefix={arXiv},
      primaryClass={cs.CV},
      url={https://arxiv.org/abs/2312.10035}, 
}

@misc{wilkinson2025contrastivelearningrobustrepresentations,
      title={Contrastive Learning for Robust Representations of Neutrino Data}, 
      author={Alex Wilkinson and Radi Radev and Saul Alonso-Monsalve},
      year={2025},
      eprint={2502.07724},
      archivePrefix={arXiv},
      primaryClass={hep-ex},
      url={https://arxiv.org/abs/2502.07724}, 
}

@article{domine2020scalable,
  title={Scalable deep convolutional neural networks for sparse, locally dense liquid argon time projection chamber data},
  author={Domin{\'e}, Laura and Terao, Kazuhiro and (DeepLearnPhysics Collaboration)},
  journal={Physical Review D},
  volume={102},
  number={1},
  pages={012005},
  year={2020},
  publisher={APS}
}

@misc{ryali2023hierahierarchicalvisiontransformer,
      title={Hiera: A Hierarchical Vision Transformer without the Bells-and-Whistles}, 
      author={Chaitanya Ryali and Yuan-Ting Hu and Daniel Bolya and Chen Wei and Haoqi Fan and Po-Yao Huang and Vaibhav Aggarwal and Arkabandhu Chowdhury and Omid Poursaeed and Judy Hoffman and Jitendra Malik and Yanghao Li and Christoph Feichtenhofer},
      year={2023},
      eprint={2306.00989},
      archivePrefix={arXiv},
      primaryClass={cs.CV},
      url={https://arxiv.org/abs/2306.00989}, 
}

@misc{loshchilov2019decoupledweightdecayregularization,
      title={Decoupled Weight Decay Regularization}, 
      author={Ilya Loshchilov and Frank Hutter},
      year={2019},
      eprint={1711.05101},
      archivePrefix={arXiv},
      primaryClass={cs.LG},
      url={https://arxiv.org/abs/1711.05101}, 
}

@article{Acciarri2018,
  title = {The Pandora multi-algorithm approach to automated pattern recognition of cosmic-ray muon and neutrino events in the MicroBooNE detector},
  volume = {78},
  ISSN = {1434-6052},
  url = {http://dx.doi.org/10.1140/epjc/s10052-017-5481-6},
  DOI = {10.1140/epjc/s10052-017-5481-6},
  number = {1},
  journal = {The European Physical Journal C},
  publisher = {Springer Science and Business Media LLC},
  author = {Acciarri,  R. and Adams,  C. and An,  R. and Anthony,  J. and Asaadi,  J. and Auger,  M. and Bagby,  L. and Balasubramanian,  S. and Baller,  B. and Barnes,  C. and Barr,  G. and Bass,  M. and Bay,  F. and Bishai,  M. and Blake,  A. and Bolton,  T. and Camilleri,  L. and Caratelli,  D. and Carls,  B. and Castillo Fernandez,  R. and Cavanna,  F. and Chen,  H. and Church,  E. and Cianci,  D. and Cohen,  E. and Collin,  G. H. and Conrad,  J. M. and Convery,  M. and Crespo-Anadón,  J. I. and Del Tutto,  M. and Devitt,  A. and Dytman,  S. and Eberly,  B. and Ereditato,  A. and Escudero Sanchez,  L. and Esquivel,  J. and Fadeeva,  A. A. and Fleming,  B. T. and Foreman,  W. and Furmanski,  A. P. and Garcia-Gamez,  D. and Garvey,  G. T. and Genty,  V. and Goeldi,  D. and Gollapinni,  S. and Graf,  N. and Gramellini,  E. and Greenlee,  H. and Grosso,  R. and Guenette,  R. and Hackenburg,  A. and Hamilton,  P. and Hen,  O. and Hewes,  J. and Hill,  C. and Ho,  J. and Horton-Smith,  G. and Hourlier,  A. and Huang,  E.-C. and James,  C. and Jan de Vries,  J. and Jen,  C.-M. and Jiang,  L. and Johnson,  R. A. and Joshi,  J. and Jostlein,  H. and Kaleko,  D. and Karagiorgi,  G. and Ketchum,  W. and Kirby,  B. and Kirby,  M. and Kobilarcik,  T. and Kreslo,  I. and Laube,  A. and Li,  Y. and Lister,  A. and Littlejohn,  B. R. and Lockwitz,  S. and Lorca,  D. and Louis,  W. C. and Luethi,  M. and Lundberg,  B. and Luo,  X. and Marchionni,  A. and Mariani,  C. and Marshall,  J. and Martinez Caicedo,  D. A. and Meddage,  V. and Miceli,  T. and Mills,  G. B. and Moon,  J. and Mooney,  M. and Moore,  C. D. and Mousseau,  J. and Murrells,  R. and Naples,  D. and Nienaber,  P. and Nowak,  J. and Palamara,  O. and Paolone,  V. and Papavassiliou,  V. and Pate,  S. F. and Pavlovic,  Z. and Piasetzky,  E. and Porzio,  D. and Pulliam,  G. and Qian,  X. and Raaf,  J. L. and Rafique,  A. and Rochester,  L. and Rudolf von Rohr,  C. and Russell,  B. and Schmitz,  D. W. and Schukraft,  A. and Seligman,  W. and Shaevitz,  M. H. and Sinclair,  J. and Smith,  A. and Snider,  E. L. and Soderberg,  M. and S\"{o}ldner-Rembold,  S. and Soleti,  S. R. and Spentzouris,  P. and Spitz,  J. and St. John,  J. and Strauss,  T. and Szelc,  A. M. and Tagg,  N. and Terao,  K. and Thomson,  M. and Toups,  M. and Tsai,  Y.-T. and Tufanli,  S. and Usher,  T. and Van De Pontseele,  W. and Van de Water,  R. G. and Viren,  B. and Weber,  M. and Wickremasinghe,  D. A. and Wolbers,  S. and Wongjirad,  T. and Woodruff,  K. and Yang,  T. and Yates,  L. and Zeller,  G. P. and Zennamo,  J. and Zhang,  C.},
  year = {2018},
  month = jan 
}

@Article{app11062455,
AUTHOR = {Majumdar, Krishanu and Mavrokoridis, Konstantinos},
TITLE = {Review of Liquid Argon Detector Technologies in the Neutrino Sector},
JOURNAL = {Applied Sciences},
VOLUME = {11},
YEAR = {2021},
NUMBER = {6},
ARTICLE-NUMBER = {2455},
URL = {https://www.mdpi.com/2076-3417/11/6/2455},
ISSN = {2076-3417},
DOI = {10.3390/app11062455}
}

@article{JMLR:v9:vandermaaten08a,
  author  = {Laurens van der Maaten and Geoffrey Hinton},
  title   = {Visualizing Data using t-SNE},
  journal = {Journal of Machine Learning Research},
  year    = {2008},
  volume  = {9},
  number  = {86},
  pages   = {2579--2605},
  url     = {http://jmlr.org/papers/v9/vandermaaten08a.html}
}

@inbook{Sudre_2017,
   title={Generalised Dice Overlap as a Deep Learning Loss Function for Highly Unbalanced Segmentations},
   ISBN={9783319675589},
   ISSN={1611-3349},
   url={http://dx.doi.org/10.1007/978-3-319-67558-9_28},
   DOI={10.1007/978-3-319-67558-9_28},
   booktitle={Deep Learning in Medical Image Analysis and Multimodal Learning for Clinical Decision Support},
   publisher={Springer International Publishing},
   author={Sudre, Carole H. and Li, Wenqi and Vercauteren, Tom and Ourselin, Sebastien and Jorge Cardoso, M.},
   year={2017},
   pages={240–248} }

@article{park2025fm4npp,
  title={FM4NPP: A Scaling Foundation Model for Nuclear and Particle Physics},
  author={Park, David and Li, Shuhang and Huang, Yi and Luo, Xihaier and Yu, Haiwang and Go, Yeonju and Pinkenburg, Christopher and Lin, Yuewei and Yoo, Shinjae and Osborn, Joseph and others},
  journal={arXiv preprint arXiv:2508.14087},
  year={2025}
}

@article{Kudryavtsev_2016,
   title={Underground physics with DUNE},
   volume={718},
   ISSN={1742-6596},
   url={http://dx.doi.org/10.1088/1742-6596/718/6/062032},
   DOI={10.1088/1742-6596/718/6/062032},
   number={6},
   journal={Journal of Physics: Conference Series},
   publisher={IOP Publishing},
   author={Kudryavtsev, Vitaly A},
   year={2016},
   month=may, pages={062032} }

@misc{polarmae,
      title={Particle Trajectory Representation Learning with Masked Point Modeling}, 
      author={Sam Young and Yeon-jae Jwa and Kazuhiro Terao},
      year={2025},
      eprint={2502.02558},
      archivePrefix={arXiv},
      primaryClass={hep-ex},
      url={https://arxiv.org/abs/2502.02558},
}

@misc{concerto,
      title={Concerto: Joint 2D-3D Self-Supervised Learning Emerges Spatial Representations}, 
      author={Yujia Zhang and Xiaoyang Wu and Yixing Lao and Chengyao Wang and Zhuotao Tian and Naiyan Wang and Hengshuang Zhao},
      year={2025},
      eprint={2510.23607},
      archivePrefix={arXiv},
      primaryClass={cs.CV},
      url={https://arxiv.org/abs/2510.23607}, 
}

@misc{huang2016deepnetworksstochasticdepth,
      title={Deep Networks with Stochastic Depth}, 
      author={Gao Huang and Yu Sun and Zhuang Liu and Daniel Sedra and Kilian Weinberger},
      year={2016},
      eprint={1603.09382},
      archivePrefix={arXiv},
      primaryClass={cs.LG},
      url={https://arxiv.org/abs/1603.09382}, 
}

@misc{touvron2021goingdeeperimagetransformers,
      title={Going deeper with Image Transformers}, 
      author={Hugo Touvron and Matthieu Cord and Alexandre Sablayrolles and Gabriel Synnaeve and Hervé Jégou},
      year={2021},
      eprint={2103.17239},
      archivePrefix={arXiv},
      primaryClass={cs.CV},
      url={https://arxiv.org/abs/2103.17239}, 
}

@misc{grill2020byol,
      title={Bootstrap your own latent: A new approach to self-supervised Learning}, 
      author={Jean-Bastien Grill and Florian Strub and Florent Altché and Corentin Tallec and Pierre H. Richemond and Elena Buchatskaya and Carl Doersch and Bernardo Avila Pires and Zhaohan Daniel Guo and Mohammad Gheshlaghi Azar and Bilal Piot and Koray Kavukcuoglu and Rémi Munos and Michal Valko},
      year={2020},
      eprint={2006.07733},
      archivePrefix={arXiv},
      primaryClass={cs.LG},
      url={https://arxiv.org/abs/2006.07733}, 
}

@misc{swav,
      title={Unsupervised Learning of Visual Features by Contrasting Cluster Assignments}, 
      author={Mathilde Caron and Ishan Misra and Julien Mairal and Priya Goyal and Piotr Bojanowski and Armand Joulin},
      year={2021},
      eprint={2006.09882},
      archivePrefix={arXiv},
      primaryClass={cs.CV},
      url={https://arxiv.org/abs/2006.09882}, 
}

@misc{leigh2024tokenizationneededmaskedparticle,
      title={Is Tokenization Needed for Masked Particle Modelling?}, 
      author={Matthew Leigh and Samuel Klein and François Charton and Tobias Golling and Lukas Heinrich and Michael Kagan and Inês Ochoa and Margarita Osadchy},
      year={2024},
      eprint={2409.12589},
      archivePrefix={arXiv},
      primaryClass={hep-ph},
      url={https://arxiv.org/abs/2409.12589}, 
}

@misc{rieck2025selfsupervisedlearningstrategiesjet,
      title={Self-Supervised Learning Strategies for Jet Physics}, 
      author={Patrick Rieck and Kyle Cranmer and Etienne Dreyer and Eilam Gross and Nilotpal Kakati and Dmitrii Kobylanskii and Garrett W. Merz and Nathalie Soybelman},
      year={2025},
      eprint={2503.11632},
      archivePrefix={arXiv},
      primaryClass={hep-ph},
      url={https://arxiv.org/abs/2503.11632}, 
}

@misc{hao2025rinorenormalizationgroupinvariance,
      title={RINO: Renormalization Group Invariance with No Labels}, 
      author={Zichun Hao and Raghav Kansal and Abhijith Gandrakota and Chang Sun and Ngadiuba Jennifer and Javier Duarte and Maria Spiropulu},
      year={2025},
      eprint={2509.07486},
      archivePrefix={arXiv},
      primaryClass={hep-ex},
      url={https://arxiv.org/abs/2509.07486}, 
}

@article{Mikuni_2025,
   title={Method to simultaneously facilitate all jet physics tasks},
   volume={111},
   ISSN={2470-0029},
   url={http://dx.doi.org/10.1103/PhysRevD.111.054015},
   DOI={10.1103/physrevd.111.054015},
   number={5},
   journal={Physical Review D},
   publisher={American Physical Society (APS)},
   author={Mikuni, Vinicius and Nachman, Benjamin},
   year={2025},
   month=mar }

@misc{bhimji2025omnilearnedfoundationmodelframework,
      title={OmniLearned: A Foundation Model Framework for All Tasks Involving Jet Physics}, 
      author={Wahid Bhimji and Chris Harris and Vinicius Mikuni and Benjamin Nachman},
      year={2025},
      eprint={2510.24066},
      archivePrefix={arXiv},
      primaryClass={hep-ph},
      url={https://arxiv.org/abs/2510.24066}, 
}

@misc{kirillov2019panopticsegmentation,
      title={Panoptic Segmentation}, 
      author={Alexander Kirillov and Kaiming He and Ross Girshick and Carsten Rother and Piotr Dollár},
      year={2019},
      eprint={1801.00868},
      archivePrefix={arXiv},
      primaryClass={cs.CV},
      url={https://arxiv.org/abs/1801.00868}, 
}

@article{Hubert1985,
  title = {Comparing partitions},
  volume = {2},
  ISSN = {1432-1343},
  url = {http://dx.doi.org/10.1007/BF01908075},
  DOI = {10.1007/bf01908075},
  number = {1},
  journal = {Journal of Classification},
  publisher = {Springer Science and Business Media LLC},
  author = {Hubert,  Lawrence and Arabie,  Phipps},
  year = {1985},
  month = dec,
  pages = {193–218}
}

@misc{detr,
      title={End-to-End Object Detection with Transformers}, 
      author={Nicolas Carion and Francisco Massa and Gabriel Synnaeve and Nicolas Usunier and Alexander Kirillov and Sergey Zagoruyko},
      year={2020},
      eprint={2005.12872},
      archivePrefix={arXiv},
      primaryClass={cs.CV},
      url={https://arxiv.org/abs/2005.12872}, 
}

@misc{mvit,
      title={Multiscale Vision Transformers}, 
      author={Haoqi Fan and Bo Xiong and Karttikeya Mangalam and Yanghao Li and Zhicheng Yan and Jitendra Malik and Christoph Feichtenhofer},
      year={2021},
      eprint={2104.11227},
      archivePrefix={arXiv},
      primaryClass={cs.CV},
      url={https://arxiv.org/abs/2104.11227}, 
}

@misc{mvitv2,
      title={MViTv2: Improved Multiscale Vision Transformers for Classification and Detection}, 
      author={Yanghao Li and Chao-Yuan Wu and Haoqi Fan and Karttikeya Mangalam and Bo Xiong and Jitendra Malik and Christoph Feichtenhofer},
      year={2022},
      eprint={2112.01526},
      archivePrefix={arXiv},
      primaryClass={cs.CV},
      url={https://arxiv.org/abs/2112.01526}, 
}

@misc{m2f,
      title={Masked-attention Mask Transformer for Universal Image Segmentation}, 
      author={Bowen Cheng and Ishan Misra and Alexander G. Schwing and Alexander Kirillov and Rohit Girdhar},
      year={2022},
      eprint={2112.01527},
      archivePrefix={arXiv},
      primaryClass={cs.CV},
      url={https://arxiv.org/abs/2112.01527}, 
}

@misc{cheng2021perpixelclassificationneedsemantic,
      title={Per-Pixel Classification is Not All You Need for Semantic Segmentation}, 
      author={Bowen Cheng and Alexander G. Schwing and Alexander Kirillov},
      year={2021},
      eprint={2107.06278},
      archivePrefix={arXiv},
      primaryClass={cs.CV},
      url={https://arxiv.org/abs/2107.06278}, 
}

@misc{kolodiazhnyi2023oneformer3dtransformerunifiedpoint,
      title={OneFormer3D: One Transformer for Unified Point Cloud Segmentation}, 
      author={Maxim Kolodiazhnyi and Anna Vorontsova and Anton Konushin and Danila Rukhovich},
      year={2023},
      eprint={2311.14405},
      archivePrefix={arXiv},
      primaryClass={cs.CV},
      url={https://arxiv.org/abs/2311.14405}, 
}

@article{Adams_2020,
   title={Reconstruction and measurement of O(100) MeV energy electromagnetic activity from pi0 arrow gamma/gamma decays in the MicroBooNE LArTPC},
   volume={15},
   ISSN={1748-0221},
   url={http://dx.doi.org/10.1088/1748-0221/15/02/P02007},
   DOI={10.1088/1748-0221/15/02/p02007},
   number={02},
   journal={Journal of Instrumentation},
   publisher={IOP Publishing},
   author={Adams, C. and Alrashed, M. and An, R. and Anthony, J. and Asaadi, J. and Ashkenazi, A. and Balasubramanian, S. and Baller, B. and Barnes, C. and Barr, G. and Basque, V. and Bass, M. and Bay, F. and Berkman, S. and Bhanderi, A. and Bhat, A. and Bishai, M. and Blake, A. and Bolton, T. and Camilleri, L. and Caratelli, D. and Terrazas, I. Caro and Carr, R. and Fernandez, R. Castillo and Cavanna, F. and Cerati, G. and Chen, Y. and Church, E. and Cianci, D. and Cohen, E.O. and Conrad, J.M. and Convery, M. and Cooper-Troendle, L. and Crespo-Anadón, J.I. and Tutto, M. Del and Devitt, D. and Diaz, A. and Domine, L. and Duffy, K. and Dytman, S. and Eberly, B. and Ereditato, A. and Sanchez, L. Escudero and Esquivel, J. and Evans, J.J. and Fitzpatrick, R.S. and Fleming, B.T. and Foppiani, N. and Franco, D. and Furmanski, A.P. and Garcia-Gamez, D. and Gardiner, S. and Genty, V. and Goeldi, D. and Gollapinni, S. and Goodwin, O. and Gramellini, E. and Green, P. and Greenlee, H. and Grosso, R. and Gu, L. and Gu, W. and Guenette, R. and Guzowski, P. and Hamilton, P. and Hen, O. and Hill, C. and Horton-Smith, G.A. and Hourlier, A. and Huang, E.-C. and Itay, R. and James, C. and de Vries, J. Jan and Ji, X. and Jiang, L. and Jo, J.H. and Johnson, R.A. and Joshi, J. and Jwa, Y.-J. and Karagiorgi, G. and Ketchum, W. and Kirby, B. and Kirby, M. and Kobilarcik, T. and Kreslo, I. and Lepetic, I. and Li, Y. and Lister, A. and Littlejohn, B.R. and Lockwitz, S. and Lorca, D. and Louis, W.C. and Luethi, M. and Lundberg, B. and Luo, X. and Marchionni, A. and Marcocci, S. and Mariani, C. and Marshall, J. and Martin-Albo, J. and Caicedo, D.A. Martinez and Mason, K. and Mastbaum, A. and McConkey, N. and Meddage, V. and Mettler, T. and Miller, K. and Mills, J. and Mistry, K. and Mogan, A. and Mohayai, T. and Moon, J. and Mooney, M. and Moore, C.D. and Mousseau, J. and Murphy, M. and Murrells, R. and Naples, D. and Neely, R.K. and Nienaber, P. and Nowak, J. and Palamara, O. and Pandey, V. and Paolone, V. and Papadopoulou, A. and Papavassiliou, V. and Pate, S.F. and Paudel, A. and Pavlovic, Z. and Piasetzky, E. and Porzio, D. and Prince, S. and Pulliam, G. and Qian, X. and Raaf, J.L. and Rafique, A. and Ren, L. and Rochester, L. and Rogers, H.E. and Ross-Lonergan, M. and Rohr, C. Rudolf von and Russell, B. and Scanavini, G. and Schmitz, D.W. and Schukraft, A. and Seligman, W. and Shaevitz, M.H. and Sharankova, R. and Sinclair, J. and Smith, A. and Snider, E.L. and Soderberg, M. and Söldner-Rembold, S. and Soleti, S.R. and Spentzouris, P. and Spitz, J. and Stancari, M. and John, J.St. and Strauss, T. and Sutton, K. and Sword-Fehlberg, S. and Szelc, A.M. and Tagg, N. and Tang, W. and Terao, K. and Thornton, R.T. and Toups, M. and Tsai, Y.-T. and Tufanli, S. and Usher, T. and Pontseele, W. Van De and de Water, R.G. Van and Viren, B. and Weber, M. and Wei, H. and Wickremasinghe, D.A. and Williams, Z. and Wolbers, S. and Wongjirad, T. and Woodruff, K. and Wospakrik, M. and Wu, W. and Yang, T. and Yarbrough, G. and Yates, L.E. and Zeller, G.P. and Zennamo, J. and Zhang, C.},
   year={2020},
   month=feb, pages={P02007–P02007} }

@article{Acciarri_2014,
   title={First Measurement of Neutrino and Antineutrino Coherent Charged Pion Production on Argon},
   volume={113},
   ISSN={1079-7114},
   url={http://dx.doi.org/10.1103/PhysRevLett.113.261801},
   DOI={10.1103/physrevlett.113.261801},
   number={26},
   journal={Physical Review Letters},
   publisher={American Physical Society (APS)},
   author={Acciarri, R. and Adams, C. and Asaadi, J. and Baller, B. and Bolton, T. and Bromberg, C. and Cavanna, F. and Church, E. and Edmunds, D. and Ereditato, A. and Farooq, S. and Fleming, B. and Greenlee, H. and Hatcher, R. and Horton-Smith, G. and James, C. and Klein, E. and Lang, K. and Laurens, P. and Mehdiyev, R. and Page, B. and Palamara, O. and Partyka, K. and Rameika, G. and Rebel, B. and Santos, E. and Schukraft, A. and Soderberg, M. and Spitz, J. and Szelc, A. M. and Weber, M. and Yang, T. and Zeller, G. P.},
   year={2014},
   month=dec }

@article{muon2021,
   title={Calorimetric classification of track-like signatures in liquid argon TPCs using MicroBooNE data},
   volume={2021},
   ISSN={1029-8479},
   url={http://dx.doi.org/10.1007/JHEP12(2021)153},
   DOI={10.1007/jhep12(2021)153},
   number={12},
   journal={Journal of High Energy Physics},
   publisher={Springer Science and Business Media LLC},
   author={Abratenko, P. and An, R. and Anthony, J. and Asaadi, J. and Ashkenazi, A. and Balasubramanian, S. and Baller, B. and Barnes, C. and Barr, G. and Basque, V. and Bathe-Peters, L. and Benevides Rodrigues, O. and Berkman, S. and Bhanderi, A. and Bhat, A. and Bishai, M. and Blake, A. and Bolton, T. and Camilleri, L. and Caratelli, D. and Caro Terrazas, I. and Castillo Fernandez, R. and Cavanna, F. and Cerati, G. and Chen, Y. and Church, E. and Cianci, D. and Conrad, J. M. and Convery, M. and Cooper-Troendle, L. and Crespo-Anadón, J. I. and Del Tutto, M. and Dennis, S. R. and Devitt, A. and Diurba, R. and Dorrill, R. and Duffy, K. and Dytman, S. and Eberly, B. and Ereditato, A. and Evans, J. J. and Fine, R. and Fiorentini Aguirre, G. A. and Fitzpatrick, R. S. and Fleming, B. T. and Foppiani, N. and Franco, D. and Furmanski, A. P. and Garcia-Gamez, D. and Gardiner, S. and Ge, G. and Gollapinni, S. and Goodwin, O. and Gramellini, E. and Green, P. and Greenlee, H. and Gu, W. and Guenette, R. and Guzowski, P. and Hagaman, L. and Hall, E. and Hamilton, P. and Hen, O. and Horton-Smith, G. A. and Hourlier, A. and Itay, R. and James, C. and Ji, X. and Jiang, L. and Jo, J. H. and Johnson, R. A. and Jwa, Y.-J. and Kamp, N. and Kaneshige, N. and Karagiorgi, G. and Ketchum, W. and Kirby, M. and Kobilarcik, T. and Kreslo, I. and LaZur, R. and Lepetic, I. and Li, K. and Li, Y. and Lin, K. and Littlejohn, B. R. and Louis, W. C. and Luo, X. and Manivannan, K. and Mariani, C. and Marsden, D. and Marshall, J. and Martinez Caicedo, D. A. and Mason, K. and Mastbaum, A. and McConkey, N. and Meddage, V. and Mettler, T. and Miller, K. and Mills, J. and Mistry, K. and Mogan, A. and Mohayai, T. and Moon, J. and Mooney, M. and Moor, A. F. and Moore, C. D. and Mora Lepin, L. and Mousseau, J. and Murphy, M. and Naples, D. and Navrer-Agasson, A. and Neely, R. K. and Nowak, J. and Nunes, M. and Palamara, O. and Paolone, V. and Papadopoulou, A. and Papavassiliou, V. and Pate, S. F. and Paudel, A. and Pavlovic, Z. and Piasetzky, E. and Ponce-Pinto, I. D. and Prince, S. and Qian, X. and Raaf, J. L. and Radeka, V. and Rafique, A. and Reggiani-Guzzo, M. and Ren, L. and Rice, L. C. J. and Rochester, L. and Rodriguez Rondon, J. and Rogers, H. E. and Rosenberg, M. and Ross-Lonergan, M. and Scanavini, G. and Schmitz, D. W. and Schukraft, A. and Seligman, W. and Shaevitz, M. H. and Sharankova, R. and Sinclair, J. and Smith, A. and Snider, E. L. and Soderberg, M. and Söldner-Rembold, S. and Spentzouris, P. and Spitz, J. and Stancari, M. and St. John, J. and Strauss, T. and Sutton, K. and Sword-Fehlberg, S. and Szelc, A. M. and Tagg, N. and Tang, W. and Terao, K. and Thorpe, C. and Totani, D. and Toups, M. and Tsai, Y.-T. and Uchida, M. A. and Usher, T. and Van De Pontseele, W. and Viren, B. and Weber, M. and Wei, H. and Williams, Z. and Wolbers, S. and Wongjirad, T. and Wospakrik, M. and Wright, N. and Wu, W. and Yandel, E. and Yang, T. and Yarbrough, G. and Yates, L. E. and Zeller, G. P. and Zennamo, J. and Zhang, C.},
   year={2021},
   month=dec }

@misc{wang2020solov2dynamicfastinstance,
      title={SOLOv2: Dynamic and Fast Instance Segmentation}, 
      author={Xinlong Wang and Rufeng Zhang and Tao Kong and Lei Li and Chunhua Shen},
      year={2020},
      eprint={2003.10152},
      archivePrefix={arXiv},
      primaryClass={cs.CV},
      url={https://arxiv.org/abs/2003.10152}, 
}

@article{Abratenko_2022,
   title={Wire-cell 3D pattern recognition techniques for neutrino event reconstruction in large LArTPCs: algorithm description and quantitative evaluation with MicroBooNE simulation},
   volume={17},
   ISSN={1748-0221},
   url={http://dx.doi.org/10.1088/1748-0221/17/01/P01037},
   DOI={10.1088/1748-0221/17/01/p01037},
   number={01},
   journal={Journal of Instrumentation},
   publisher={IOP Publishing},
   author={Abratenko, P. and An, R. and Anthony, J. and Arellano, L. and Asaadi, J. and Ashkenazi, A. and Balasubramanian, S. and Baller, B. and Barnes, C. and Barr, G. and Basque, V. and Bathe-Peters, L. and Benevides Rodrigues, O. and Berkman, S. and Bhanderi, A. and Bhat, A. and Bishai, M. and Blake, A. and Bolton, T. and Book, J.Y. and Camilleri, L. and Caratelli, D. and Caro Terrazas, I. and Castillo Fernandez, R. and Cavanna, F. and Cerati, G. and Chen, Y. and Cianci, D. and Conrad, J.M. and Convery, M. and Cooper-Troendle, L. and Crespo-Anadón, J.I. and Del Tutto, M. and Dennis, S.R. and Detje, P. and Devitt, A. and Diurba, R. and Dorrill, R. and Duffy, K. and Dytman, S. and Eberly, B. and Ereditato, A. and Evans, J.J. and Fine, R. and Fiorentini Aguirre, G.A. and Fitzpatrick, R.S. and Fleming, B.T. and Foppiani, N. and Franco, D. and Furmanski, A.P. and Garcia-Gamez, D. and Gardiner, S. and Ge, G. and Gollapinni, S. and Goodwin, O. and Gramellini, E. and Green, P. and Greenlee, H. and Gu, W. and Guenette, R. and Guzowski, P. and Hagaman, L. and Hen, O. and Hilgenberg, C. and Horton-Smith, G.A. and Hourlier, A. and Itay, R. and James, C. and Ji, X. and Jiang, L. and Jo, J.H. and Johnson, R.A. and Jwa, Y.-J. and Kalra, D. and Kamp, N. and Kaneshige, N. and Karagiorgi, G. and Ketchum, W. and Kirby, M. and Kobilarcik, T. and Kreslo, I. and LaZur, R. and Lepetic, I. and Li, K. and Li, Y. and Lin, K. and Littlejohn, B.R. and Louis, W.C. and Luo, X. and Manivannan, K. and Mariani, C. and Marsden, D. and Marshall, J. and Martinez Caicedo, D.A. and Mason, K. and Mastbaum, A. and McConkey, N. and Meddage, V. and Mettler, T. and Miller, K. and Mills, J. and Mistry, K. and Mogan, A. and Mohayai, T. and Moon, J. and Mooney, M. and Moor, A.F. and Moore, C.D. and Mora Lepin, L. and Mousseau, J. and Murphy, M. and Naples, D. and Navrer-Agasson, A. and Nebot-Guinot, M. and Neely, R.K. and Newmark, D.A. and Nowak, J. and Nunes, M. and Palamara, O. and Paolone, V. and Papadopoulou, A. and Papavassiliou, V. and Pate, S.F. and Patel, N. and Paudel, A. and Pavlovic, Z. and Piasetzky, E. and Ponce-Pinto, I.D. and Prince, S. and Qian, X. and Raaf, J.L. and Radeka, V. and Rafique, A. and Reggiani-Guzzo, M. and Ren, L. and Rice, L.C.J. and Rochester, L. and Rodriguez Rondon, J. and Rosenberg, M. and Ross-Lonergan, M. and Scanavini, G. and Schmitz, D.W. and Schukraft, A. and Seligman, W. and Shaevitz, M.H. and Sharankova, R. and Shi, J. and Sinclair, J. and Smith, A. and Snider, E.L. and Soderberg, M. and Söldner-Rembold, S. and Spentzouris, P. and Spitz, J. and Stancari, M. and St. John, J. and Strauss, T. and Sutton, K. and Sword-Fehlberg, S. and Szelc, A.M. and Tagg, N. and Tang, W. and Terao, K. and Thorpe, C. and Totani, D. and Toups, M. and Tsai, Y.-T. and Uchida, M.A. and Usher, T. and Van De Pontseele, W. and Viren, B. and Weber, M. and Wei, H. and Williams, Z. and Wolbers, S. and Wongjirad, T. and Wospakrik, M. and Wresilo, K. and Wright, N. and Wu, W. and Yandel, E. and Yang, T. and Yarbrough, G. and Yates, L.E. and Yu, H.W. and Zeller, G.P. and Zennamo, J. and Zhang, C.},
   year={2022},
   month=jan, pages={P01037} }

@misc{gu2022efficientlymodelinglongsequences,
      title={Efficiently Modeling Long Sequences with Structured State Spaces}, 
      author={Albert Gu and Karan Goel and Christopher Ré},
      year={2022},
      eprint={2111.00396},
      archivePrefix={arXiv},
      primaryClass={cs.LG},
      url={https://arxiv.org/abs/2111.00396}, 
}

@misc{gu2024mambalineartimesequencemodeling,
      title={Mamba: Linear-Time Sequence Modeling with Selective State Spaces}, 
      author={Albert Gu and Tri Dao},
      year={2024},
      eprint={2312.00752},
      archivePrefix={arXiv},
      primaryClass={cs.LG},
      url={https://arxiv.org/abs/2312.00752}, 
}

@misc{salimans2016weightnormalizationsimplereparameterization,
      title={Weight Normalization: A Simple Reparameterization to Accelerate Training of Deep Neural Networks}, 
      author={Tim Salimans and Diederik P. Kingma},
      year={2016},
      eprint={1602.07868},
      archivePrefix={arXiv},
      primaryClass={cs.LG},
      url={https://arxiv.org/abs/1602.07868}, 
}

@misc{koh2020scalableproposalfreeinstancesegmentation,
      title={Scalable, Proposal-free Instance Segmentation Network for 3D Pixel Clustering and Particle Trajectory Reconstruction in Liquid Argon Time Projection Chambers}, 
      author={Dae Heun Koh and Pierre Côte de Soux and Laura Dominé and François Drielsma and Ran Itay and Qing Lin and Kazuhiro Terao and Ka Vang Tsang and Tracy Usher},
      year={2020},
      eprint={2007.03083},
      archivePrefix={arXiv},
      primaryClass={physics.ins-det},
      url={https://arxiv.org/abs/2007.03083}, 
}

@article{Heggestuen_2024,
doi = {10.1088/1748-0221/19/05/C05038},
url = {https://doi.org/10.1088/1748-0221/19/05/C05038},
year = {2024},
month = {may},
publisher = {IOP Publishing},
volume = {19},
number = {05},
pages = {C05038},
author = {Heggestuen, A. and on behalf of the ICARUS collaboration},
title = {Light detection and cosmic rejection in the ICARUS LArTPC at Fermilab},
journal = {Journal of Instrumentation}
}

@article{Foreman_2020,
   title={Calorimetry for low-energy electrons using charge and light in liquid argon},
   volume={101},
   ISSN={2470-0029},
   url={http://dx.doi.org/10.1103/PhysRevD.101.012010},
   DOI={10.1103/physrevd.101.012010},
   number={1},
   journal={Physical Review D},
   publisher={American Physical Society (APS)},
   author={Foreman, W. and Acciarri, R. and Asaadi, J. A. and Badgett, W. and Blaszczyk, F. d. M. and Bouabid, R. and Bromberg, C. and Carey, R. and Cavanna, F. and Cevallos Aleman, J. I. and Chatterjee, A. and Evans, J. and Falcone, A. and Flanagan, W. and Fleming, B. T. and Garcia-Gamez, D. and Gelli, B. and Ghosh, T. and Gomes, R. A. and Gramellini, E. and Gran, R. and Hamilton, P. and Hill, C. and Ho, J. and Hugon, J. and Iwai, E. and Kearns, E. and Kemp, E. and Kobilarcik, T. and Kordosky, M. and Kryczyński, P. and Lang, K. and Linehan, R. and Machado, A. A. B. and Maruyama, T. and Metcalf, W. and Moura, C. A. and Nichol, R. and Nunes, M. and Nutini, I. and Olivier, A. and Palamara, O. and Paley, J. and Paulucci, L. and Pulliam, G. and Raaf, J. L. and Rebel, B. and Rodrigues, O. and Mendes Santos, L. and Schmitz, D. W. and Segreto, E. and Smith, D. and Soderberg, M. and Spagliardi, F. and John, J. M. St. and Stancari, M. and Szelc, A. M. and Tzanov, M. and Walker, D. and Williams, Z. and Yang, T. and Yu, J. and Zhang, S.},
   year={2020},
   month=jan }

@article{Adams_2020_0,
   title={Calibration of the charge and energy loss per unit length of the MicroBooNE liquid argon time projection chamber using muons and protons},
   volume={15},
   ISSN={1748-0221},
   url={http://dx.doi.org/10.1088/1748-0221/15/03/P03022},
   DOI={10.1088/1748-0221/15/03/p03022},
   number={03},
   journal={Journal of Instrumentation},
   publisher={IOP Publishing},
   author={Adams, C. and Alrashed, M. and An, R. and Anthony, J. and Asaadi, J. and Ashkenazi, A. and Balasubramanian, S. and Baller, B. and Barnes, C. and Barr, G. and Basque, V. and Bass, M. and Bay, F. and Berkman, S. and Bhanderi, A. and Bhat, A. and Bishai, M. and Blake, A. and Bolton, T. and Camilleri, L. and Caratelli, D. and Terrazas, I. Caro and Carr, R. and Fernandez, R. Castillo and Cavanna, F. and Cerati, G. and Chen, Y. and Church, E. and Cianci, D. and Cohen, E.O. and Conrad, J.M. and Convery, M. and Cooper-Troendle, L. and Crespo-Anadón, J.I. and Tutto, M. Del and Devitt, D. and Diaz, A. and Domine, L. and Duffy, K. and Dytman, S. and Eberly, B. and Ereditato, A. and Sanchez, L. Escudero and Esquivel, J. and Evans, J.J. and Fitzpatrick, R.S. and Fleming, B.T. and Foppiani, N. and Franco, D. and Furmanski, A.P. and Garcia-Gamez, D. and Gardiner, S. and Genty, V. and Goeldi, D. and Gollapinni, S. and Goodwin, O. and Gramellini, E. and Green, P. and Greenlee, H. and Grosso, R. and Gu, L. and Gu, W. and Guenette, R. and Guzowski, P. and Hamilton, P. and Hen, O. and Hill, C. and Horton-Smith, G.A. and Hourlier, A. and Huang, E.-C. and Itay, R. and James, C. and de Vries, J. Jan and Ji, X. and Jiang, L. and Jo, J.H. and Johnson, R.A. and Joshi, J. and Jwa, Y.-J. and Karagiorgi, G. and Ketchum, W. and Kirby, B. and Kirby, M. and Kobilarcik, T. and Kreslo, I. and Lepetic, I. and Li, Y. and Lister, A. and Littlejohn, B.R. and Lockwitz, S. and Lorca, D. and Louis, W.C. and Luethi, M. and Lundberg, B. and Luo, X. and Marchionni, A. and Marcocci, S. and Mariani, C. and Marshall, J. and Martin-Albo, J. and Caicedo, D.A. Martinez and Mason, K. and Mastbaum, A. and McConkey, N. and Meddage, V. and Mettler, T. and Miller, K. and Mills, J. and Mistry, K. and Mogan, A. and Mohayai, T. and Moon, J. and Mooney, M. and Moore, C.D. and Mousseau, J. and Murphy, M. and Murrells, R. and Naples, D. and Neely, R.K. and Nienaber, P. and Nowak, J. and Palamara, O. and Pandey, V. and Paolone, V. and Papadopoulou, A. and Papavassiliou, V. and Pate, S.F. and Paudel, A. and Pavlovic, Z. and Piasetzky, E. and Porzio, D. and Prince, S. and Pulliam, G. and Qian, X. and Raaf, J.L. and Rafique, A. and Ren, L. and Rochester, L. and Rogers, H.E. and Ross-Lonergan, M. and Rohr, C. Rudolf von and Russell, B. and Scanavini, G. and Schmitz, D.W. and Schukraft, A. and Seligman, W. and Shaevitz, M.H. and Sharankova, R. and Sinclair, J. and Smith, A. and Snider, E.L. and Soderberg, M. and Söldner-Rembold, S. and Soleti, S.R. and Spentzouris, P. and Spitz, J. and Stancari, M. and John, J. St. and Strauss, T. and Sutton, K. and Sword-Fehlberg, S. and Szelc, A.M. and Tagg, N. and Tang, W. and Terao, K. and Thornton, R.T. and Toups, M. and Tsai, Y.-T. and Tufanli, S. and Usher, T. and Pontseele, W. Van De and de Water, R.G. Van and Viren, B. and Weber, M. and Wei, H. and Wickremasinghe, D.A. and Williams, Z. and Wolbers, S. and Wongjirad, T. and Woodruff, K. and Wospakrik, M. and Wu, W. and Yang, T. and Yarbrough, G. and Yates, L.E. and Zeller, G.P. and Zennamo, J. and Zhang, C.},
   year={2020},
   month=mar, pages={P03022–P03022} }

@article{Acciarri_2017_0,
   title={Michel electron reconstruction using cosmic-ray data from the MicroBooNE LArTPC},
   volume={12},
   ISSN={1748-0221},
   url={http://dx.doi.org/10.1088/1748-0221/12/09/P09014},
   DOI={10.1088/1748-0221/12/09/p09014},
   number={09},
   journal={Journal of Instrumentation},
   publisher={IOP Publishing},
   author={Acciarri, R. and Adams, C. and An, R. and Anthony, J. and Asaadi, J. and Auger, M. and Bagby, L. and Balasubramanian, S. and Baller, B. and Barnes, C. and Barr, G. and Bass, M. and Bay, F. and Bishai, M. and Blake, A. and Bolton, T. and Bugel, L. and Camilleri, L. and Caratelli, D. and Carls, B. and Fernandez, R. Castillo and Cavanna, F. and Chen, H. and Church, E. and Cianci, D. and Cohen, E. and Collin, G.H. and Conrad, J.M. and Convery, M. and Crespo-Anadón, J.I. and Tutto, M. Del and Devitt, D. and Dytman, S. and Eberly, B. and Ereditato, A. and Sanchez, L. Escudero and Esquivel, J. and Fleming, B.T. and Foreman, W. and Furmanski, A.P. and Garcia-Gamez, D. and Garvey, G.T. and Genty, V. and Goeldi, D. and Gollapinni, S. and Graf, N. and Gramellini, E. and Greenlee, H. and Grosso, R. and Guenette, R. and Hackenburg, A. and Hamilton, P. and Hen, O. and Hewes, J. and Hill, C. and Ho, J. and Horton-Smith, G. and Huang, E.-C. and James, C. and de Vries, J. Jan and Jen, C.-M. and Jiang, L. and Johnson, R.A. and Joshi, J. and Jostlein, H. and Kaleko, D. and Karagiorgi, G. and Ketchum, W. and Kirby, B. and Kirby, M. and Kobilarcik, T. and Kreslo, I. and Laube, A. and Li, Y. and Lister, A. and Littlejohn, B.R. and Lockwitz, S. and Lorca, D. and Louis, W.C. and Luethi, M. and Lundberg, B. and Luo, X. and Marchionni, A. and Mariani, C. and Marshall, J. and Caicedo, D.A. Martinez and Meddage, V. and Miceli, T. and Mills, G.B. and Moon, J. and Mooney, M. and Moore, C.D. and Mousseau, J. and Murrells, R. and Naples, D. and Nienaber, P. and Nowak, J. and Palamara, O. and Paolone, V. and Papavassiliou, V. and Pate, S.F. and Pavlovic, Z. and Piasetzky, E. and Porzio, D. and Pulliam, G. and Qian, X. and Raaf, J.L. and Rafique, A. and Rochester, L. and Rohr, C. Rudolf von and Russell, B. and Schmitz, D.W. and Schukraft, A. and Seligman, W. and Shaevitz, M.H. and Sinclair, J. and Snider, E.L. and Soderberg, M. and Söldner-Rembold, S. and Soleti, S.R. and Spentzouris, P. and Spitz, J. and John, J. St. and Strauss, T. and Sutton, K.A. and Szelc, A.M. and Tagg, N. and Terao, K. and Thomson, M. and Toups, M. and Tsai, Y.-T. and Tufanli, S. and Usher, T. and de Water, R.G. Van and Viren, B. and Weber, M. and Wickremasinghe, D.A. and Wolbers, S. and Wongjirad, T. and Woodruff, K. and Yang, T. and Yates, L. and Zeller, G.P. and Zennamo, J. and Zhang, C.},
   year={2017},
   month=sep, pages={P09014–P09014} }

@article{Kaleko_2013,
   title={PMT triggering and readout for the MicroBooNE experiment},
   volume={8},
   ISSN={1748-0221},
   url={http://dx.doi.org/10.1088/1748-0221/8/09/C09009},
   DOI={10.1088/1748-0221/8/09/c09009},
   number={09},
   journal={Journal of Instrumentation},
   publisher={IOP Publishing},
   author={Kaleko, D},
   year={2013},
   month=sep, pages={C09009–C09009} }

@misc{poli2023hyenahierarchylargerconvolutional,
      title={Hyena Hierarchy: Towards Larger Convolutional Language Models}, 
      author={Michael Poli and Stefano Massaroli and Eric Nguyen and Daniel Y. Fu and Tri Dao and Stephen Baccus and Yoshua Bengio and Stefano Ermon and Christopher Ré},
      year={2023},
      eprint={2302.10866},
      archivePrefix={arXiv},
      primaryClass={cs.LG},
      url={https://arxiv.org/abs/2302.10866}, 
}

@misc{ku2025systemsalgorithmsconvolutionalmultihybrid,
      title={Systems and Algorithms for Convolutional Multi-Hybrid Language Models at Scale}, 
      author={Jerome Ku and Eric Nguyen and David W. Romero and Garyk Brixi and Brandon Yang and Anton Vorontsov and Ali Taghibakhshi and Amy X. Lu and Dave P. Burke and Greg Brockman and Stefano Massaroli and Christopher Ré and Patrick D. Hsu and Brian L. Hie and Stefano Ermon and Michael Poli},
      year={2025},
      eprint={2503.01868},
      archivePrefix={arXiv},
      primaryClass={cs.LG},
      url={https://arxiv.org/abs/2503.01868}, 
}

@article{Qian_2018,
   title={Three-dimensional imaging for large LArTPCs},
   volume={13},
   ISSN={1748-0221},
   url={http://dx.doi.org/10.1088/1748-0221/13/05/P05032},
   DOI={10.1088/1748-0221/13/05/p05032},
   number={05},
   journal={Journal of Instrumentation},
   publisher={IOP Publishing},
   author={Qian, X. and Zhang, C. and Viren, B. and Diwan, M.},
   year={2018},
   month=may, pages={P05032–P05032} }

@article{larpix,
   title={LArPix: demonstration of low-power 3D pixelated charge readout for liquid argon time projection chambers},
   volume={13},
   ISSN={1748-0221},
   url={http://dx.doi.org/10.1088/1748-0221/13/10/P10007},
   DOI={10.1088/1748-0221/13/10/p10007},
   number={10},
   journal={Journal of Instrumentation},
   publisher={IOP Publishing},
   author={Dwyer, D.A. and Garcia-Sciveres, M. and Gnani, D. and Grace, C. and Kohn, S. and Kramer, M. and Krieger, A. and Lin, C.J. and Luk, K.B. and Madigan, P. and Marshall, C. and Steiner, H. and Stezelberger, T.},
   year={2018},
   month=oct, pages={P10007–P10007} }

@inproceedings{lee2015deeply,
  title={Deeply-supervised nets},
  author={Lee, Chen-Yu and Xie, Saining and Gallagher, Patrick and Zhang, Zhengyou and Tu, Zhuowen},
  booktitle={Artificial intelligence and statistics},
  pages={562--570},
  year={2015},
  organization={Pmlr}
}
